\documentclass[twocolumn]{aastex701}

\usepackage{amsmath}
\usepackage[normalem]{ulem}
\usepackage{xcolor}
\usepackage{etoolbox} 
\usepackage{physics}
\usepackage{url}

\newcommand{\lya}{Ly$\alpha$}

\newcommand{\HI}{H\,\textsc{i}}

\newcommand{\hcd}{HCD}
\newcommand{\hcds}{HCDs}
\newcommand{\dla}{DLA}
\newcommand{\dlas}{DLAs}
\newcommand{\lls}{LLS}
\newcommand{\llss}{LLSs}
\newcommand{\poneD}{\ensuremath{P_{\rm 1D}}}

\newcommand{\NHI}{\ensuremath{N_{\rm HI}}}
\newcommand{\Rvir}{\ensuremath{R_{200\mathrm{c}}}}
\newcommand{\Mvir}{\ensuremath{M_{200\mathrm{c}}}}

\newcommand{\kms}{\ensuremath{\mathrm{km\,s^{-1}}}}
\newcommand{\Msun}{\ensuremath{M_\odot}}
\newcommand{\hcell}{\ell}

\newif\ifshowcomments
\showcommentstrue

\definecolor{todored}{HTML}{C0392B}

\definecolor{old}{HTML}{C0392B}
\definecolor{new}{HTML}{21866D}

\shorttitle{A halo-based framework for HCDs}
\shortauthors{Soares da Silva et al.}

\submitjournal{ApJ}

\graphicspath{{./}{figures/}{plots/}}

\begin{document}

\title{Where the Forest Goes Dark: Halo-centered Characterization of High Column Density Systems with IllustrisTNG}

\correspondingauthor{Victor R. Soares da Silva}

\author[orcid=0000-0002-2060-2290,gname=Victor R.,sname={Soares da Silva}]{Victor R. Soares da Silva}
\affiliation{Department of Mathematical Physics, Institute of Physics, University of S\~ao Paulo, R. do Mat\~ao 1371, 05508-090, S\~ao Paulo, SP, Brazil}
\email[show]{victorroberto132@usp.br}

\author[orcid=0000-0002-0279-4874,gname=Daniel,sname={López-Cano}]{Daniel López-Cano}
\affiliation{Department of Mathematical Physics, Institute of Physics, University of S\~ao Paulo, R. do Mat\~ao 1371, 05508-090, S\~ao Paulo, SP, Brazil}
\email{daniellopezcano13@gmail.com}

\author[orcid=0000-0001-8295-7022,gname={L. Raul},sname=Abramo]{L. Raul Abramo}
\affiliation{Department of Mathematical Physics, Institute of Physics, University of S\~ao Paulo, R. do Mat\~ao 1371, 05508-090, S\~ao Paulo, SP, Brazil}
\email{raulabramo@usp.br}

\author[orcid=0000-0002-9553-4261,gname={Jon\'as},sname={Chaves-Montero}]{J. Chaves-Montero}
\affiliation{Institut de F\'{i}sica d’Altes Energies (IFAE), The Barcelona Institute of Science and Technology, Edifici Cn, Campus UAB, 08193, Bellaterra (Barcelona), Spain }
\email{jchaves@ifae.es}

\author[orcid=0000-0002-0937-0644,gname={Francisco},sname=Maion]{Francisco Maion}
\affiliation{Columbia Astrophysics Laboratory, Columbia University, 550 West 120th Street, New York, NY 10027, USA}
\email{fgmaion@gmail.com}

\begin{abstract} High column density systems of neutral hydrogen (HCDs) --- Lyman limit systems (LLSs), sub-damped and damped Ly$\alpha$ absorbers (DLAs) --- contaminate both the 3D and the 1D statistics of the Ly$\alpha$ forest. Most DLAs are masked out from the analyses, but cleaning algorithms are not perfect, and weaker LLSs are individually undetectable, so the residual contamination is absorbed into nuisance parameters that dilute the cosmological constraining power. Simulations provide the ideal setting to study this effect, allowing absorption features to be directly traced back to the gas producing them. We identify high-column-density structures on the gas cells of the TNG50 hydrodynamical simulation, deblend them in velocity space, and split every sightline into HCD-only and forest-only spectra. Applied to halos at $z \simeq 3$ out to 50 virial radii, it yields each class's covering fraction against impact parameter $b$ and halo mass. The impact parameter defines the type of absorption appearing in the spectrum: DLAs give way to sub-DLAs at $b \approx 0.11\,R_{200c}$, sub-DLAs to LLSs at $0.3$, and LLSs to the Ly$\alpha$-forest at $0.5$, which covers $\simeq 78\%$ of sightlines at $R_{200c}$. In units of $R_{200c}$ the sequence is nearly mass-independent over three decades in mass, so the virial radius sets the scale of the neutral gas distribution. A single Voigt component recovers the column density of essentially every damped system, but not of LLSs, whose absorption features often arise from several separate contributions from gas structures along the sightline. Our detailed characterization of HCDs gives the first step to the construction of advanced techniques to directly forward-model their contribution to Ly$\alpha$ spectra. \end{abstract}


\section{Introduction}
\label{sec:intro}

The \lya\ forest --- the dense set of absorption features imprinted on the spectra of high-redshift quasars (QSOs) by intergalactic neutral hydrogen \citep[for reviews, see][]{Rauch1998,Meiksin2009,McQuinn2016} --- is one of the few probes of structure formation available at $2 \lesssim z \lesssim 5$. Because the transmitted flux traces the underlying density field \citep[e.g.,][]{Cen1994,Hernquist1996,Croft1998}, its statistics constrain the matter power spectrum on quasi-linear and non-linear scales \citep[e.g.,][]{McDonald2006}, the thermal and ionization state of the intergalactic medium \citep[IGM;][]{HuiGnedin1997,Schaye2000,Walther2019}, and the expansion history through baryon acoustic oscillations \citep[BAO;][]{Busca2013,Bautista2017,duMasdesBourboux2020}.

The number of available forests has grown by more than two orders of magnitude
in two decades: from samples of order $10^{3}$ spectra in the early 2000s
\citep{McDonald2006} to $\simeq820\,000$ in the second data release of
the Dark Energy Spectroscopic Instrument \citep[DESI;][]{DESI2016a,DESI2022instr,DESI2025DR2Lya}.
These data constrain the isotropic BAO scale at $z_{\rm eff}=2.33$ to $0.65\%$
and the line-of-sight scale, which fixes $H(z)$, to $1.1\%$
\citep{DESI2025DR2Lya}, and support full-shape analyses of the
three-dimensional correlations that extract the Alcock--Paczy\'nski signal in addition to the acoustic scale \citep{2026arXiv260727410D}.
The first data release already yielded measurements of the one-dimensional flux
power spectrum (\poneD) with two independent estimators
\citep{2025JCAP...10..004K,2025JCAP...11..079R}, extending the BOSS and eBOSS
sequence \citep{Palanque_Delabrouille_2013,Chabanier_2019,Ravoux_2023} and
constraining the amplitude and logarithmic slope of the linear matter power
spectrum at $z=3$ \citep{2026arXiv260121432C}.

Statistical uncertainties play a central role in these measurements and the treatment of astrophysical contaminants  costs a substantial part of the constraining
power.
Among the leading contaminants are high column density systems (\hcds):
regions of neutral hydrogen dense enough to be optically thick,
encountered when a sightline passes through
the circumgalactic medium (CGM) of a galaxy or other dense gas.
The absorption features of these systems saturate the transmitted
flux over several \AA{} of spectrum, where the forest features it hides are
themselves a fraction of an \AA{} wide. The affected range grows
with column density.
\hcds\ are classified by their \HI column density, \NHI, into Lyman limit systems (\llss), sub-damped
\lya\ systems (sub-\dlas) and damped \lya\ systems (\dlas), in order of
increasing self-shielding.\footnote{Class boundaries in $\log_{10}(\NHI/\mathrm{cm^{-2}})$, in order of
increasing column density: the diffuse forest below $17.2$; \llss\ between
$17.2$ and $19$, optically thick to ionizing radiation; sub-\dlas\ between $19$
and $20.3$, partially self-shielded; and \dlas\ above $20.3$, essentially fully
self-shielded \citep{2005ARA&A..43..861W}. We split the damped regime further at
$21.0$ into small and large \dlas, since the extent of the contamination varies
substantially across that range.}

The resulting contamination of the forest observables is scale dependent, and
has been measured. \citet{2012JCAP...07..028F} showed analytically (and also employing mock
spectra) that unmasked \hcds\ raise the effective \lya\ bias and lower the
redshift-space distortion parameter, and \citet{2011MNRAS.415.2257M} estimated
the \dla\ correlations between sightlines.
\citet{2018MNRAS.474.3032R} and
\citet{2018MNRAS.476.3716R} measured the effect in cosmological hydrodynamical
simulations, provided multiplicative \poneD\ templates as a function of \NHI,
$k_\parallel$ and redshift, and found that residual contamination of the
three-dimensional flux power spectrum reaches the ten-percent level at
$k \sim 0.1\,h\,\mathrm{Mpc}^{-1}$ even after the strongest systems are masked.

Survey analyses adopt these prescriptions. BAO and full-shape fits marginalize
over an \hcd\ bias and a wing-scale parameter
\citep{duMasdesBourboux2020,Adame_2025}. The DESI DR1 \poneD\ analysis splits
the population into column-density groups following
\citet{2018MNRAS.474.3032R}, because \dla\ finders lose completeness rapidly
below the damped threshold \citep{2026arXiv260121432C}. The amplitude of each
group is left free during inference, but that freedom is expensive: The \lls\ amplitudes rank among the nuisance
parameters most strongly correlated with the recovered cosmology, marginalizing over them inflates the uncertainty on the compressed parameters by ${\simeq}50\%$, and dropping the \lls\ and sub-\dla\ terms shifts the inferred parameters by more than $3\sigma$ \citep{2026arXiv260121432C}.
\citet{2025JCAP...11..074T} tested a Voigt-based alternative against mocks with
known \hcd\ content and found it comparable to the templates in use.

What all of these descriptions have in common is that they parametrize \hcds\ by
their column density and by an effective large-scale bias, calibrated either on
the observed column-density distribution function or on the measured
\dla--forest cross-correlation
\citep{2012JCAP...11..059F,2018MNRAS.473.3019P}. The link to the halos that host
the absorbers has been studied mostly through this integrated bias, or through
the abundance and metallicity of the \dla\ population
\citep{2014MNRAS.445.2313B}, rather than through the spectral signature that a
sightline actually records as a function of where it passes relative to a halo.
Observations cannot supply this
description. A spectrum records the absorption but not the gas that produced it,
so a feature can be attributed to a halo of known mass at a known transverse
distance only in simulations. A forward model that places \hcds\ on a simulated
density or halo field, rather than drawing them from a global distribution,
requires exactly that attribution, and we measure it here.
\citet{2025arXiv250708940M} takes the complementary route in TNG50, starting from
strong blended \lya\ absorbers, at $\log\NHI\simeq16$ and therefore below the
\lls\ threshold, and identifying the halos they trace; we start from the halos
and measure what each sightline records as a function of its distance from them.

\hcds\ are also of interest in their own right. Their incidence, kinematics and line profiles depend on the state of the CGM gas and the dense IGM, and therefore on self-shielding, accretion and feedback \citep{Schaye2001,2013MNRAS.430.2427R,peroux2024multiscalemultiphasecircumgalacticmedium}. A characterization organized by halo properties is thus useful on both fronts.

\begin{figure}[t!]
    \centering
    \includegraphics[width=\columnwidth]{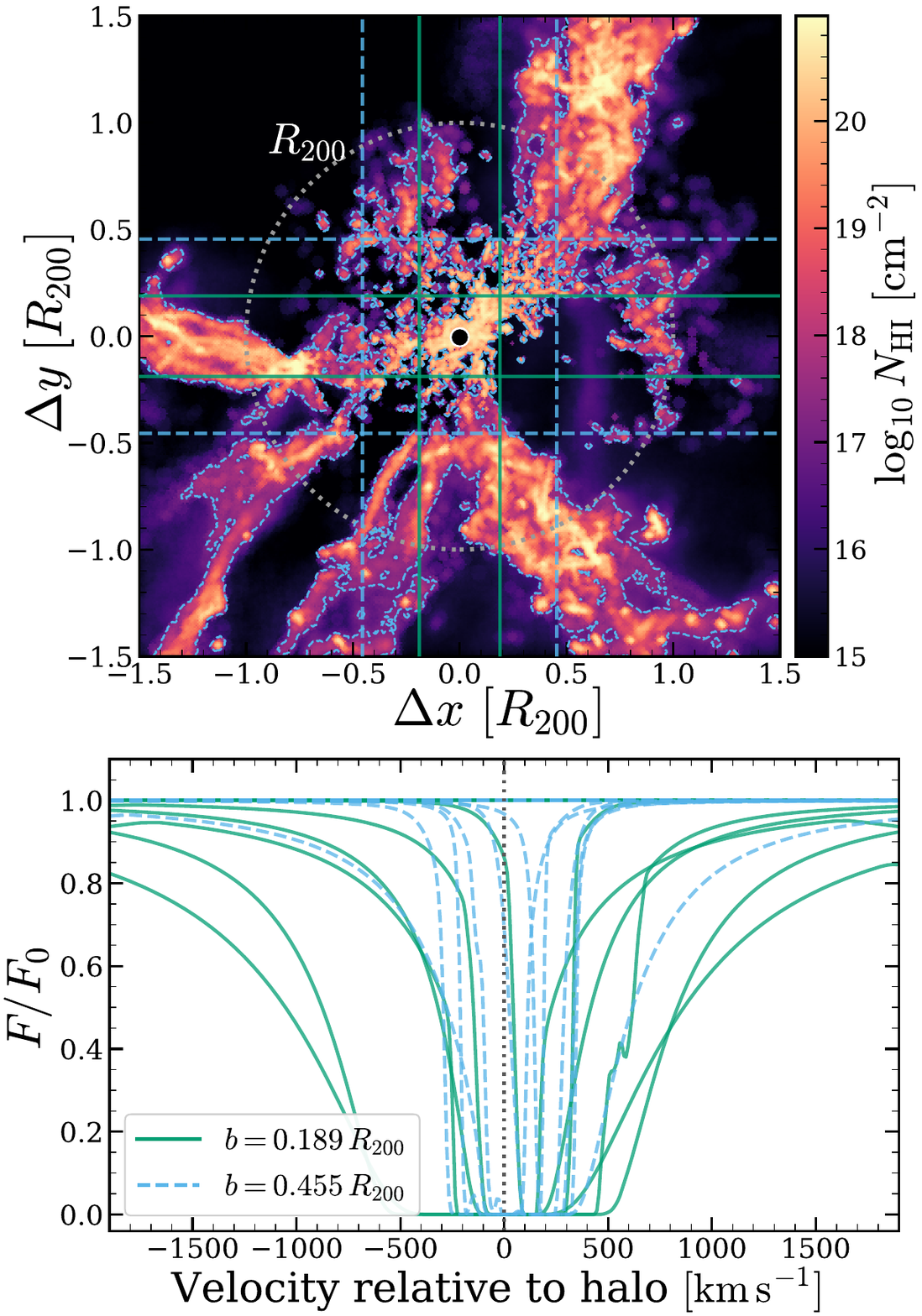}
\caption{The halo-centered sampling of this work, for a halo of
$\Mvir \approx 10^{11.86}\,\Msun$ at $z\simeq3$. \textbf{Top:} \HI\ column
density projected along the $z$ axis, with the halo centre marked by the black
dot and \Rvir\ by the grey dotted circle. The blue dashed contour traces
$\NHI = 10^{17.2}\,\mathrm{cm^{-2}}$, the \hcd\ threshold of
Sect.~\ref{subsec:hcd_isolation}. Straight lines mark the sightlines at two
impact parameters, $b = 0.189\,\Rvir$ (green, solid) and $b = 0.455\,\Rvir$
(blue, dashed); those along $x$ appear as horizontal lines and those along $y$
as vertical lines, while the four along $z$ are seen end-on and are not marked.
\textbf{Bottom:} \texttt{HCD\_only} transmitted flux recorded by the twelve
sightlines at each impact parameter, including those along $z$, against velocity
relative to the systemic velocity of the halo, with the same colour and
line-style coding.}
    \label{fig:halo_skewers_example}
\end{figure}

In this work we measure the imprint of \hcds\ on synthetic \lya\ skewers
crossing the CGM and near environment of halos at $z \simeq 3$.
Figure~\ref{fig:halo_skewers_example} illustrates the approach on a single
halo: sightlines are placed at controlled distances from a halo of known mass,
and the absorption each one records is measured as a function of that distance.

Our measurements rest on three ingredients.
First, we extend the \texttt{fake\_spectra} package \citep{2017_Bird_theory_Lya_mock_Fake_Spectra} with a skewer-level identification procedure that flags the gas cells belonging to \hcd\ structures before computing the optical depth, separates structures whose absorption an observer could resolve into distinct systems, and returns two complementary spectra per sightline: \texttt{HCD\_only}, carrying
the flagged structures, and \texttt{Forest\_only},
carrying the absorption that remains once they are removed.\footnote{\url{https://github.com/VictorRoberto132/FakeSpectra_DLAs}
}
Second, we design a halo-centered sampling campaign that pierces individual haloes along a controlled grid of impact parameters, so that the resulting absorption statistics can be read directly as functions of halo mass and halo-centric distance. Third, we run the entire campaign on three levels of the TNG50 resolution ladder \citep{2017MNRAS.465.3291W,2018MNRAS.475..676S,2018MNRAS.473.4077P,2018MNRAS.477.1206N,2018MNRAS.475..648P,2019ComAC...6....2N}, which share a single baryonic model and therefore isolate numerical resolution as a systematic to be quantified rather than as a physical effect.

These three ingredients yield a systematic characterization of \hcd\ absorption
as a function of halo properties: how the absorber population is organized
radially around halos, how that organization depends on host-halo mass, and how
far a single-absorber description holds for
the systems the pipeline isolates.

The paper is organized as follows. Section~\ref{sec:simulations_skewers}
describes the simulations, the skewer-extraction pipeline and the \hcd\ isolation
procedure, and defines the halo-centered sampling campaign.
Section~\ref{sec:results} presents the measurements: the radial organization of
the absorber population, its dependence on host-halo mass and the skewer-level
column-density distributions (Section~\ref{subsec:nhi_statistics}), and the
Voigt-profile characterization (Section~\ref{subsec:voigt_diagnostics}).
Section~\ref{sec:discussion} summarizes our conclusions and their implications
for forest modeling. Appendix~\ref{app:resolution_tests} establishes the
numerical convergence of every observable used in the main text and defines the
range of impact parameters over which we regard our measurements reliable.

Throughout, we adopt the cosmology of the IllustrisTNG simulations,
$\Omega_{\rm m}=0.3089$, $\Omega_{\rm b}=0.0486$, $\Omega_\Lambda=0.6911$,
$h=0.6774$, $\sigma_8=0.8159$ and $n_{\rm s}=0.9667$
\citep{2016A&A...594A..13P}.

\section{Methods}\label{sec:simulations_skewers}

In the following subsections we explain how this work generates synthetic \lya\ spectra from cosmological hydrodynamical simulations and isolate \hcd\ features. Working with simulations grants direct access to the gas
distribution behind every absorption feature, so each absorption can be assigned
to the structure that produces it. This section describes the
simulations (Sect.~\ref{subsec:simulation_suites}), the extraction of the
skewers and the construction of the spectra
(Sect.~\ref{subsec:fake_spectra_skewers}), the method we have devised for isolating the \hcd\ contributions (Sect.~\ref{subsec:hcd_isolation}), and the halo sample the absorbers are
associated with (Sect.~\ref{subsec:halo_sample_skewers}).
Throughout this work we denote with $b$ the impact parameter of a given sightline with respect to the corresponding halo center, and reserve $b_{\rm D}$ for the Doppler parameter.

\subsection{The IllustrisTNG simulations}\label{subsec:simulation_suites}

We employ the IllustrisTNG suite \citep{Pillepich_2019,2019ComAC...6....2N}, run
with the moving-mesh code \textsc{arepo} \citep{Springel_2010}, which solves the
magnetohydrodynamic equations on an unstructured Voronoi tessellation that moves
with the flow. TNG50 is the smallest and best resolved of its three volumes,
$35\,h^{-1}\,\mathrm{cMpc}$ on a side, and has been run at four resolution
levels, TNG50-1 to TNG50-4, separated by factors of $8$ in mass. The four levels share the same galaxy-formation model.
Table~\ref{tab:tng50} lists the parameters relevant to this work.

\begin{deluxetable}{lcccc}
\tabletypesize{\footnotesize}
\tablecolumns{5}
\tablewidth{\columnwidth}
\tablecaption{Numerical parameters of the TNG50 runs used in this work. All runs have a box side of $35\,h^{-1}\,\mathrm{cMpc}$
($51.7\,\mathrm{cMpc}$) and share the same galaxy-formation model. $m_{\rm gas}$
is the target gas cell mass; individual cell masses vary around it within a
factor of two by construction.
$d_{\rm gas}\equiv L_{\rm box}/N_{\rm gas}^{1/3}$ is the mean gas cell spacing. Along the text we employ $h=0.6774$.\label{tab:tng50}}
\tablehead{
\colhead{Run} & \colhead{$N_{\rm gas}$} & \colhead{$m_{\rm gas}$ [\Msun]} &
\colhead{$m_{\rm DM}$ [\Msun]} & \colhead{$d_{\rm gas}$ [$h^{-1}$ckpc]}
}
\startdata
TNG50-1\tablenotemark{a} & $2160^3$ & $8.5\times10^{4}$ & $4.5\times10^{5}$ & $16$ \\
TNG50-2 & $1080^3$ & $6.8\times10^{5}$ & $3.6\times10^{6}$ & $32$ \\
TNG50-3 & $540^3$  & $5.4\times10^{6}$ & $2.9\times10^{7}$ & $65$ \\
TNG50-4 & $270^3$  & $4.3\times10^{7}$ & $2.3\times10^{8}$ & $130$ \\
\enddata
\tablenotetext{a}{Not used in this work; listed for reference in the convergence
discussion of Sect.~\ref{subsec:simulation_suites}.}
\end{deluxetable}

Three ingredients of that model act directly on the absorbers we measure. Gas
denser than $n_{\rm H}=0.106\,\mathrm{cm^{-3}}$
is placed on the two-phase
effective equation of state of \citet{2003MNRAS.339..289S},
so its thermal state is not resolved and its neutral fraction has to be set in post-processing (Sect.~\ref{subsec:fake_spectra_skewers}). The cooling uses the \citet{2009ApJ...703.1416F} ultraviolet background with the
self-shielding prescription of \citet{2013MNRAS.430.2427R}, which sets the hydrogen density above which the gas self-shields and becomes largely neutral. Galactic winds \citep{2018MNRAS.473.4077P} and low-state kinetic AGN feedback \citep{2017MNRAS.465.3291W} set how much neutral gas the circumgalactic medium retains and how it is distributed, so the cross-sections we report inherit their calibration.

We run the pipeline of
Sects.~\ref{subsec:fake_spectra_skewers}--\ref{subsec:halo_sample_skewers} on
TNG50-4, TNG50-3 and TNG50-2, and adopt TNG50-2 as the production run for all
results quoted in the main text. The three levels share an identical baryonic
model and differ only in the number of resolution elements, so any difference
between them measures resolution alone. Appendix~\ref{app:resolution_tests}
repeats every measurement of this paper along the three levels and sets the
range of impact parameters over which our results are converged.

Convergence studies of the diffuse \lya\ forest require cell sizes of order
$20\,h^{-1}\,\mathrm{ckpc}$ for percent-level flux statistics
\citep{2015MNRAS.446.3697L,2017MNRAS.464..897B}, which the larger volumes of the
suite do not reach: TNG100-1 and TNG300-1 sample their boxes at $41\,h^{-1}\,\mathrm{ckpc}$ and
$82\,h^{-1}\,\mathrm{ckpc}$, against $32\,h^{-1}\,\mathrm{ckpc}$ for TNG50-2.
This criterion is calibrated on the underdense gas that dominates the
transmission, and is not the one that governs the observable of this paper,
which is the absorption produced by self-shielded, overdense gas.
Two published criteria and the trend along our own ladder
indicate that TNG50-2 already meets the requirement of the problem.
\citet{2014MNRAS.445.2313B} recover the DLA cross-section--halo mass relation in
a $25\,h^{-1}\,\mathrm{cMpc}$ box sampled at a mean interparticle spacing of
$49\,h^{-1}\,\mathrm{ckpc}$, against a higher-resolution companion run; TNG50-2,
at $32\,h^{-1}\,\mathrm{ckpc}$, is finer than that reference.
A second reference point is the mass-resolution criterion of
\citet{2009MNRAS.398L..26B}, calibrated on the convergence of the flux
statistics. The requirement they find becomes stricter with increasing redshift,
from $m_{\rm gas}\lesssim1.6\times10^{6}\,h^{-1}\,\Msun$
at $z=2$ to
$2\times10^{5}\,h^{-1}\,\Msun$ at $z=5$, so the latter is a conservative bound at
the $z\simeq3$ of our analysis. TNG50-2, at
$m_{\rm gas}=4.6\times10^{5}\,h^{-1}\,\Msun$
(Table~\ref{tab:tng50}, converted to $h^{-1}\Msun$), sits within a factor of ${\sim}2$ of
that bound.
In any case, the convergence of the specific measurements of this paper is
established internally: Appendix~\ref{app:resolution_tests} compares the three
levels for every observable we report, and the differences shrink from the
TNG50-4 $\rightarrow$ TNG50-3 step to the TNG50-3 $\rightarrow$ TNG50-2 one. We use these comparisons to assess the impact of numerical resolution and define the validity range of our measurements.

The campaign is designed to separate two sources of variation: halo mass,
which sets the depth of the potential well and the reservoir of neutral gas, and 
impact parameter; a third
axis, the sensitivity to the adopted baryonic-physics model, is left to future
work using the zoom-in resimulations of \citet{2026arXiv260713151M}.

\subsection{Synthetic \lya\ skewers with \texttt{fake\_spectra}}
\label{subsec:fake_spectra_skewers}

We generate synthetic spectra with \texttt{fake\_spectra}
\citep{2017_Bird_theory_Lya_mock_Fake_Spectra}, which reads the gas cells of an
\textsc{arepo} or \textsc{gadget} snapshot, selects those intersecting a given
sightline, and deposits their neutral hydrogen density, temperature and
line-of-sight peculiar velocity on a one-dimensional grid. As is standard in
forest work, that grid is in velocity units, so that positions and peculiar
velocities enter the line profile in the same units. The snapshot is treated as
a single redshift, and a cell at comoving distance $\chi$ along the sightline is
assigned the Hubble velocity $u_{\rm H}=a\,H(z_\mathrm{snap})\,\chi$, periodic on
$[0,\,aH(z_\mathrm{snap})L_{\rm box})$. The optical depth at velocity $u$ is
\begin{align}\label{eq:tau}
    \tau(u) = \sigma_\alpha\,c \int n_{\rm HI}(l)\,
    \mathcal{V}\!\left(u - u_{\rm H}(l) - v_{\rm pec}(l),\, b_{\rm D}(l)\right)
    \,\dd{l},
\end{align}
integrated over proper path length $l$ along the sightline, with $n_{\rm HI}$
[cm$^{-3}$] the proper neutral hydrogen number density, $v_{\rm pec}$ [\kms] the
line-of-sight peculiar velocity, and
$\sigma_\alpha=\sqrt{3\pi\sigma_{\rm T}/8}\,\lambda_0 f$ [cm$^{2}$] the
integrated line strength for a transition of rest-frame wavelength $\lambda_0$
and oscillator strength $f$. $\mathcal{V}$ is the Voigt profile normalized to
unit integral in velocity [s\,km$^{-1}$], with thermal Doppler parameter
$b_{\rm D}=\sqrt{2k_{\rm B}T/m_{\rm H}}$ [\kms]. The optical-depth value
assigned to each pixel is Eq.~\eqref{eq:tau} averaged over the pixel width.

The neutral fraction of star-forming cells is not resolved by the simulation and is assigned in post-processing. For cells below the star-formation threshold, $n_{\rm H} < 0.106\,\mathrm{cm^{-3}}$, the neutral fraction follows from the ionization equilibrium computed during the run, including the self-shielding attenuation of the photoionization rate described in Sect.~\ref{subsec:simulation_suites}; \texttt{fake\_spectra} re-evaluates that same \citet{2013MNRAS.430.2427R} correction from the density and temperature stored in the snapshot. Cells on the effective equation of state require a separate treatment: the neutral fraction recorded in the snapshot for them is evaluated at the mass-weighted effective temperature of the two-phase medium, a quantity that describes neither of the two phases and that underestimates the neutral content of the cell \citep{2018ApJ...866..135V}. We adopt the default prescription of \texttt{fake\_spectra} and take star-forming gas to be fully neutral, which is the limit of the \citet{2003MNRAS.339..289S} model in which the cold clouds carry essentially all of the mass and are themselves neutral.\footnote{The cold-phase mass fraction of the model is ${\simeq}90\%$ over the density range relevant here, so this assumption overestimates $n_{\rm HI}$ in star-forming cells by ${\lesssim}0.05$~dex. Such cells lie in the saturated cores of the strongest systems, where the transmitted flux is zero irrespective of this choice and the column density is set by the damping wings, so the effect on the quantities we measure is negligible.} No further self-shielding correction is applied on top of this, since gas at these densities is already deep in the self-shielded regime.

We compute the optical depth of the \HI\ \lya\ transition
alone. Sightlines run along one of the
three principal axes of the box and request a pixel width of $\Delta v = 10\;\kms$,
which the code rounds to an integer number of pixels: at $z\simeq3$ the box
spans $a H(z_\mathrm{snap}) L_{\rm box} = 3961.5\;\kms$, giving $396$ pixels. This is below the thermal width of $10^{4}\,\mathrm{K}$ gas, $b_{\rm D}^{\rm th}=12.8\;\kms$,
so the sampling does not limit the structure a
spectrum can carry, and it is finer than the resolution of the surveys the
measurements are meant to inform,\footnote{DESI observes the forest at $R\simeq2000$--$3000$
\citep{DESI2022instr}, corresponding to $c/R\simeq100$--$150\;\kms$; WEAVE-QSO
observes at $R\simeq5000$ in its wide survey and $R\simeq20\,000$ in its
high-resolution mode \citep{2016sf2a.conf..259P}, corresponding to
$\simeq60$ and $\simeq15\;\kms$ respectively.} so the
skewers can be degraded to any instrumental configuration without being
regenerated.
The spectral grid inherits the periodicity of the box. Cells within the
deposition radius of a boundary contribute at both ends of the skewer, and
absorptions pushed past $v_{\rm max}$ by peculiar velocities re-enter at $v=0$, and conversely.
The effect is negligible for the forest, but not for damped systems, whose
Lorentzian wings extend over hundreds of \kms\ and can appear split between the
two ends of a skewer.
The integral of Eq.~\eqref{eq:tau} runs over the whole box; no aperture is
imposed at this stage (see
Sect.~\ref{subsec:halo_sample_skewers}).

\textsc{arepo} discretises the gas into Voronoi cells of approximately uniform
internal density, so a cell carries no smoothing kernel. We therefore deposit
each cell as a uniform sphere of radius $\hcell=V^{1/3}$, with $V$ the cell
volume, so that a cell at perpendicular distance $d_\perp$ from the sightline
contributes the chord
\begin{align}\label{eq:tophat}
    N = \frac{3}{4\pi}\,n_{\rm HI}\,2\sqrt{\hcell^{2}-d_\perp^{2}}
\end{align}
to the column density. The factor $3/4\pi$ compensates for the choice
$\hcell=V^{1/3}$ rather than the equal-volume radius, so that the deposition
conserves the neutral hydrogen content $n_{\rm HI}V$ of the cell exactly.

\subsection{Isolating the \hcd\ contribution}
\label{subsec:hcd_isolation}

We extend \texttt{fake\_spectra} with two complementary extraction modes,
\texttt{HCD\_only} and \texttt{Forest\_only}, which decompose the absorption
along a sightline into the contribution of high-column-density structures and
that of the diffuse forest; the statistics of Sect.~\ref{sec:results} are
measured on the former, and the latter is the absorption the same sightline
would show in their absence.

The decomposition is applied to the gas cells rather than to the flux. In
observations an \hcd\ is identified from the shape it imprints on the spectrum,
so the selection is necessarily performed at the pixel level. In a simulation
the contribution of each cell is known individually, which lets the selection
act on the cells themselves and guarantees the two channels are complementary
by construction. A flux-level cut is not viable in any case: the damping wings
of an \hcd\ extend well beyond the pixels containing its saturated core, so
masking those pixels alone would leave the wings in the forest channel while
discarding part of the \hcd\ signal. The selection is therefore applied to the
cells, prior to the interpolation of Eq.~\eqref{eq:tau}.

Assignment of cells to structures is performed in velocity space. At the redshifts of
interest the peculiar velocity term of gas cells frequently exceeds the Hubble term, so  gas widely
separated in $\chi$ can be projected into the trough of an \hcd, contributing to
its absorption equivalent width without being separable from it. Velocity space
is therefore the frame in which the criteria below are defined.
The algorithm developed here for isolating the contribution from \hcds\ comprises six steps (and is illustrated in Fig.~\ref{fig:hcd_detection}):

\begin{figure}
    \centering
    \includegraphics[width=\columnwidth]{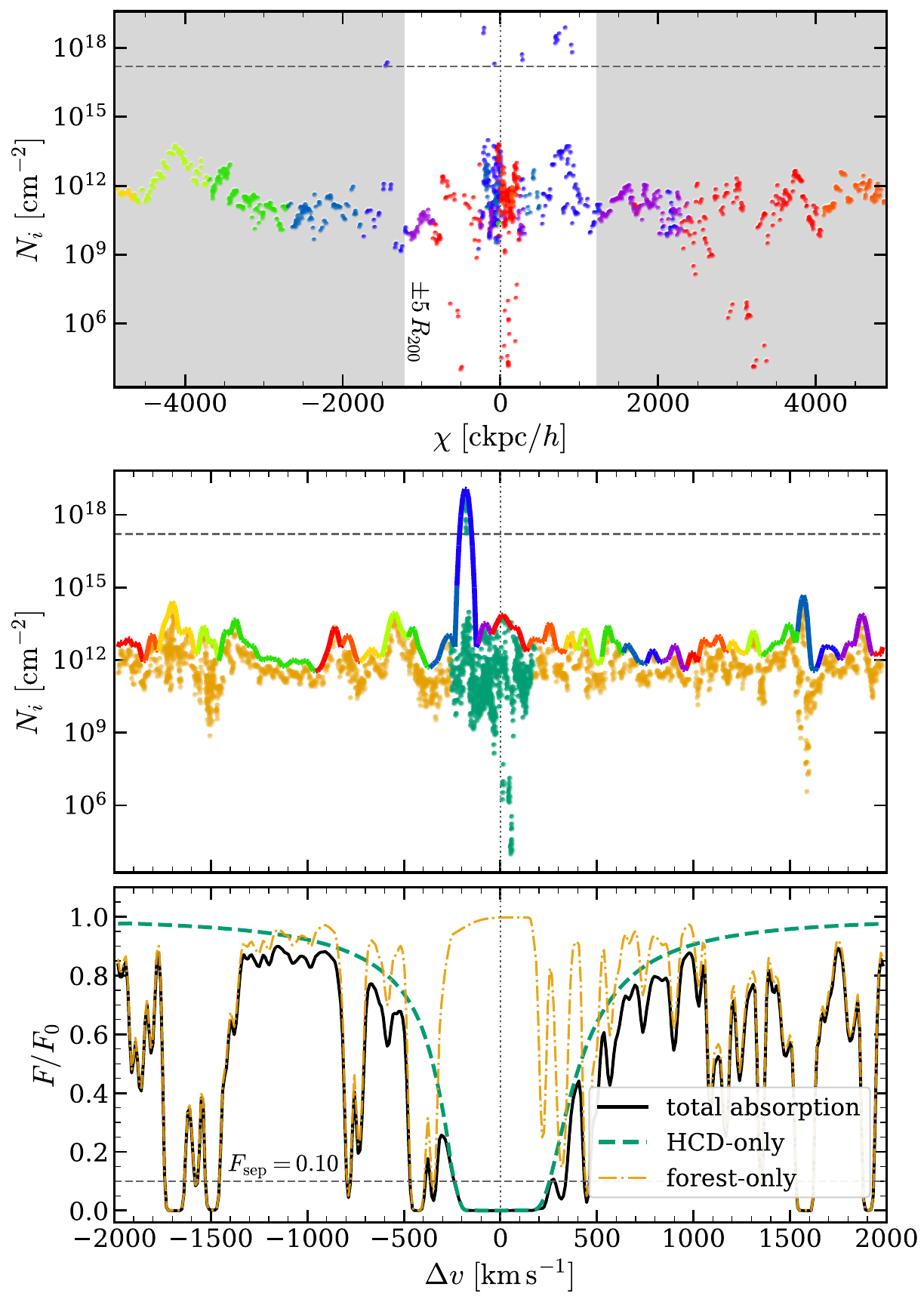}
\caption{The \hcd\ identification procedure applied to a single sightline.
\textbf{Top}: column density $N_i$ contributed by each cell intersecting the
sightline, against comoving position $\chi$, coloured by the basin it is
assigned to in velocity space by the segmentation of steps~2--3, here carried
back to configuration space. Grey bands mark $|\chi|>5\,\Rvir$ from the halo
centre, outside the association window of step~4; the dashed line marks
$10^{17.2}\,\mathrm{cm^{-2}}$ for reference.
{\textbf{Middle}:} the same cells projected onto the velocity axis (step~2), together with the smoothed profile on which
the watershed operates. The profile is coloured by the basins of step~3, using
the same colours as the top panel; the cells are coloured instead by their final
assignment to the \texttt{HCD\_only} (green) or \texttt{Forest\_only} (orange)
channel (step~5), matching the bottom panel.
{\textbf{Bottom}:} total absorption and the resulting \texttt{HCD\_only} and
\texttt{Forest\_only} spectra (step~6), with the dashed line marking the
separation level $F_{\rm sep}=0.10$ of step~5.}
    \label{fig:hcd_detection}
\end{figure}

\begin{enumerate}

\item \textbf{Contributed column per cell.} For each cell intersecting the
sightline we evaluate the column it deposits --- the top-hat
kernel of Sect.~\ref{subsec:fake_spectra_skewers} (top panel of Fig.~\ref{fig:hcd_detection}). This is the
quantity accumulated by the interpolator into the column-density array,
$\sum_i N_i$ reproducing the code's total to machine precision, and is directly
comparable with the $10^{17.2}\,\mathrm{cm^{-2}}$ threshold.

\item \textbf{Projection onto the velocity axis.} Each cell is assigned the
velocity $u_i = aH(z_\mathrm{snap})\chi_i + v_{{\rm pec},i}$ and its column
deposited in the corresponding pixel employing a nearest-grid-point scheme (middle-panel scatter), giving the
redshift-space \NHI\ profile of the sightline. The
profile is afterwards convolved with a Gaussian of width
$\sigma=b_{\rm D}^{\rm th}/\sqrt{2}={9.1}\;\kms$
corresponding to the thermal width of
$T=10^{4}\,\mathrm{K}$ gas,\footnote{We adopt a single width, set by the temperature of
the photoionized \lya\ forest rather than by the temperature of each cell,
because the quantity being imposed is a resolution floor: two components closer
than the thermal width of forest gas cannot be separated by any line-fitting
procedure applied to a real spectrum, whatever the temperature of the gas that
produced them.} returning the smooth curve over the scatter points in the middle panel of Fig~\ref{fig:hcd_detection}.

\item \textbf{Segmentation.} The smoothed profile is partitioned into basins by
a periodic one-dimensional watershed, with boundaries at its local minima.
Adjacent maxima are retained as distinct only if the intervening minimum lies
below a fraction $R=0.5$ --- a dimensionless contrast ratio of
column densities on the smoothed profile --- of the weaker of the two; shallower minima are merged,
preventing internal substructure from fragmenting a single absorber. Each cell is
assigned to the basin containing its velocity, and each basin carries
$N_b=\sum_{i\in b}N_i$, summed over cells rather than inferred from a profile
fit.\footnote{Summing $N_i$ over cells sharing a basin is a bookkeeping step,
not a statement that their absorption blends in the flux: optical depth is not
additive in the way column density is, and two structures placed in the same
basin can still leave a recovering flux between them. The physically relevant
blending is checked separately, on the flux, in step~5; the smoothing scale
adopted in step~2 matches the thermal width of the forest gas responsible for
that blending, so the two criteria are consistent with each other.} This
segmentation colours the smoothed curve in the middle panel, which in turn is employed to colour the corresponding points associated with each basin in the top panel; the underlying
cells in the middle panel (scatter points) are coloured instead by the outcome of step~5, shared with the
bottom-panel spectra.

\item \textbf{Halo association.} A basin is retained as a candidate if the cell
dominating its column lies within $\pm5\,\Rvir$ of the halo centre along the
sightline --- the only criterion evaluated in configuration space. Basins
failing it are assigned to the forest channel irrespective of their column
density. This associates already-identified structures with the halo; it does
not restrict the volume from which cells are drawn, since the optical depth of
Eq.~\eqref{eq:tau} is always integrated across the full periodic box. The window
must be wide enough to contain the gas the halo governs and narrow enough to
exclude unrelated structures along the sightline. Appendix~\ref{app:aperture}
shows that the covering fractions are insensitive to this choice for
$b\lesssim0.5\,\Rvir$, and that beyond $b\simeq1\,\Rvir$ removing the aperture
recovers the incidence of a random sightline, which is how the large-$b$
behaviour of Sect.~\ref{subsec:nhi_statistics} should be read.

\item \textbf{Classification and blending.} Candidates with $N_b\ge10^{17.2}\,\mathrm{cm^{-2}}$ 
are accepted as \hcd-like systems. Sub-threshold candidates are then tested sequentially 
in decreasing order of $N_b$. For each, we combine its optical depth with the current 
accepted set and evaluate the maximum dimensionless transmittance, $F_{\rm max} = F/F_0$, 
between their absorption minima. If $F_{\rm max} < F_{\rm sep} = 0.1$, the candidate is 
blended and merged into the accepted set; otherwise, it is assigned to the forest channel. We use transmittance rather than optical depth because, even though $\tau$
 can drop from $10^6$ to $20$ between saturated components, the transmittance remains effectively zero. Physically, the strict $F_{\rm sep}=0.1$ threshold merges sufficiently nearby forest components into LLSs when their contribution is observationally indistinguishable (it is not possible to deblend it at the level of the absorption profile). For DLAs, it cleanly deblends distant forest lines where the flux recovers, preventing observationally distinct absorbers from being artificially swallowed by the DLA's broad wings.

\item \textbf{Masking.} A boolean mask flags cells belonging to retained basins.
In \texttt{HCD\_only} mode only these cells enter the optical depth calculation (Eq.~\ref{eq:tau}),
producing strong systems with their damping wings; in \texttt{Forest\_only}
mode they are excluded and the remaining cells employed for the computation instead. Because
each cell belongs to exactly one basin, the partition is exact: the two
channels' optical depths add and their fluxes multiply, recovering the total
by construction. As shown in the bottom panel of Fig.~\ref{fig:hcd_detection}.

\end{enumerate}

The procedure rests on four parameters. One is physical: the threshold
$10^{17.2}\,\mathrm{cm^{-2}}$ of step~5, which defines the absorber class and is
the standard \hcd\ boundary. The other three are choices internal to the method:
the smoothing width $\sigma$ of step~2, fixed by the thermal broadening of
forest gas; the dimensionless contrast ratio $R$ of step~3, which sets how deep a
minimum must be for two maxima to count as separate structures; and the
separation level $F_{\rm sep}$ of step~5, the transmittance at which two
absorption features become observationally distinct. Step~4 adds a fifth choice,
the $\pm5\,\Rvir$ association window, which does not affect the identification of
structures but decides which of them are credited to the target halo.
Appendix~\ref{app:aperture} quantifies the dependence of our measurements on
that window, comparing the fiducial aperture against $\pm10\,\Rvir$ and against
no aperture at all.

\subsection{The halo-centered skewer campaign}\label{subsec:halo_sample_skewers}

Characterizing \hcd\ absorption as a function of halo properties requires
sightlines placed at controlled distances from known halos rather than drawn at
random through the box. This subsection defines that sampling, whose
construction is visually depicted in Fig.~\ref{fig:halo_skewers_example}.

\textbf{Halo selection.} 
Halos are taken from the friends-of-friends (FoF) group catalogue of the
$z\simeq3$ snapshot of TNG50-2 and characterized by their spherical-overdensity
mass and radius, $\Mvir$ [\Msun] and $\Rvir$.
We select a total of $637$ halos in five
logarithmic mass bins with edges
$\log_{10}(\Mvir/\Msun) = [10.0,\,10.5,\,11.0,\,11.5,\,12.0,\,13.0]$, drawing
$150$ halos uniformly in each of the four lower bins, and taking all $37$ available in the highest.

\textbf{Impact-parameter grid.} Each halo is pierced at its center and at $30$
logarithmically spaced impact parameters, $b$, from $0.01\,\Rvir$, deep in the inner
CGM, to $50\,\Rvir$, in the surrounding large-scale environment. We quote $b$ in
units of \Rvir, which places halos of different mass on a common footing. At $b=0$ we extract one skewer along
each principal axis of the box. At every non-zero $b$ we extract $12$ sightlines: for each
of the three axes, four skewers displaced above, below, left and right of the
halo center. This gives $3 + 30\times12 = 363$
skewers per halo and $637\times363\simeq2.3\times10^{5}$ per resolution level.

\textbf{Halo centring.} Profile widths can only be meaningfully compared in a defined rest
frame, which we set in two steps. We first shift the velocity axis so that $v=0$
is the systemic velocity of the halo, the local Hubble flow at its center of
mass plus its bulk peculiar velocity, wrapping periodically and interpolating
onto a fine uniform grid. Internal gas motions then offset the deepest
absorption from that velocity by tens of \kms, so we re-center each retained
system on the flux minimum of its own profile. Which feature to center on
is fixed by the segmentation of Sect.~\ref{subsec:hcd_isolation}.

\textbf{Recorded quantities.} For every skewer, we store the host halo properties 
(\Mvir, \Rvir), the skewer's density and temperature profiles using the \texttt{HCD\_only} mode, and the column density \NHI\ of each retained 
system. Sightlines hosting multiple systems are labeled by the total summed \NHI.

We fit a single-component Voigt profile to each system, mirroring observational 
analyses. The fit yields the inferred column density $\NHI^{\rm fit}$ and a Doppler 
parameter $b_{\rm D}^{\rm fit}$, which we convert to an effective temperature 
$T^{\rm fit} = m_{\rm H}(b_{\rm D}^{\rm fit})^{2}/2k_{\rm B}$. The residual between 
this fit and the \texttt{HCD\_only} flux quantifies how much the system departs from 
a single absorbing element (e.g., due to multiple components or complex internal 
structure). Section~\ref{subsec:voigt_diagnostics} analyzes the origin of these departures.

\textbf{Error estimation.} We repeat the halo selection five times, each with its
own random seed, and run the full campaign on each realization. Throughout this
work, every measurement is then reported as the mean of the five realizations,
drawn as a solid line, and a band of $\pm3\sigma$ about it, with $\sigma$ the
standard deviation across the five. The band is deliberately conservative: with
five realizations $\sigma$ is itself a noisy estimator, so a narrower interval
would understate the uncertainty. What it measures is the sample variance of the
halo population, not the Poisson noise of individual skewers.

Halos are drawn without replacement within a realization but independently
across realizations. In the four lower mass bins the $150$ halos drawn are a
small fraction of those available, so the realizations overlap. The highest-mass bin is exhaustive and therefore
identical in all five realizations, so no sample-variance estimate is available
for it and no band is drawn.

\textbf{Resolution floor on the impact parameter.} The impact parameter grid extends inwards beyond the scale on which the measurements are reliable, so we define a floor $b_{\rm min}=0.1\,\Rvir$ below which the results are shown but not interpreted. The limit is not set by the resolution of the gas: \textsc{arepo} refines quasi-Lagrangian, so the cells are smallest exactly where the sightlines pass closest to the halo center, and the transverse offset between adjacent impact parameters on our logarithmic grid, $\Delta b\simeq0.31\,b$, shrinks together with them, remaining well below the local cell size at every $b$ we sample. It is set instead by the collisionless component, since the gas responds to a potential dominated by dark matter. Inside the convergence radius of \citet{2003MNRAS.338...14P}, where two-body relaxation acts over the dynamical time of the halo, that potential is artificially softened and the gas distribution it supports is unreliable however finely the cells resolve it. This radius scales with the number of enclosed particles, so $r_{\rm min}/\Rvir$ is largest in the lowest mass bin, which we adopt as the worst case: at its lower edge, $\Mvir=10^{10}\,\Msun$, the dark-matter particle mass of TNG50-2 (Table~\ref{tab:tng50}) together with the cosmic dark-matter fraction gives $N(<\Rvir)\approx2\times10^{3}$, for which the Power et al.\ criterion places the reliable radius at a few to ${\sim}10\%$ of $\Rvir$, consistent with the $b_{\rm min}=0.1\,\Rvir$ found empirically from the level-to-level convergence of Appendix~\ref{app:resolution_tests}, so the two independent estimates agree. Besides resolution, the innermost sightlines cross star-forming gas whose neutral fraction is set by the prescription of Sect.~\ref{subsec:fake_spectra_skewers} rather than by the hydrodynamics; this is a limitation of the model, not of the sampling, and it is not removed at higher resolution. Appendix~\ref{app:resolution_tests} gives $r_{\rm min}$ per mass bin and compares it with the level-to-level convergence of the observables. The floor is marked as a shaded region in every radial figure.

\section{Results}
\label{sec:results}

We apply the pipeline of Sect.~\ref{sec:simulations_skewers} to characterize
\hcd\ absorption around halos. Sect.~\ref{subsec:nhi_statistics} uses the column
density measured on each sightline to establish how the absorber population is
organized radially; Sect.~\ref{subsec:voigt_diagnostics} examines the shape of
the absorption profile and what the single-component Voigt description captures.
All results are quoted for TNG50-2 at $z\simeq3$. Following
Sect.~\ref{subsec:halo_sample_skewers}.

\subsection{The radial organization of the absorber population}
\label{subsec:nhi_statistics}

\subsubsection{Incidence of absorber classes}
\label{subsubsec:covering_fraction}

We characterize the distribution of neutral gas around halos through the covering
fraction $f_C(b)$: at each impact parameter it is the
fraction of sightlines whose \texttt{HCD\_only} column density falls in one of
the five classes of Sect.~\ref{sec:intro}, with boundaries at
$\log_{10}(\NHI/\mathrm{cm^{-2}}) = 17.2$, $19$, $20.3$ and $21.0$ separating the
diffuse forest, \llss, sub-\dlas, and small and large \dlas\ in order of
increasing column density. Sightlines carrying more than one retained system are
classified by the summed \NHI\ (Sect.~\ref{subsec:halo_sample_skewers}). The
intervals partition the full range of \NHI, so the five curves sum to unity at
every $b$ by construction. $f_C(b)$ is therefore a decomposition of the sightline
population, not the cumulative covering fraction $f_C(>\NHI)$ usually reported.

\begin{figure}
    \centering
    \includegraphics[width=\columnwidth]{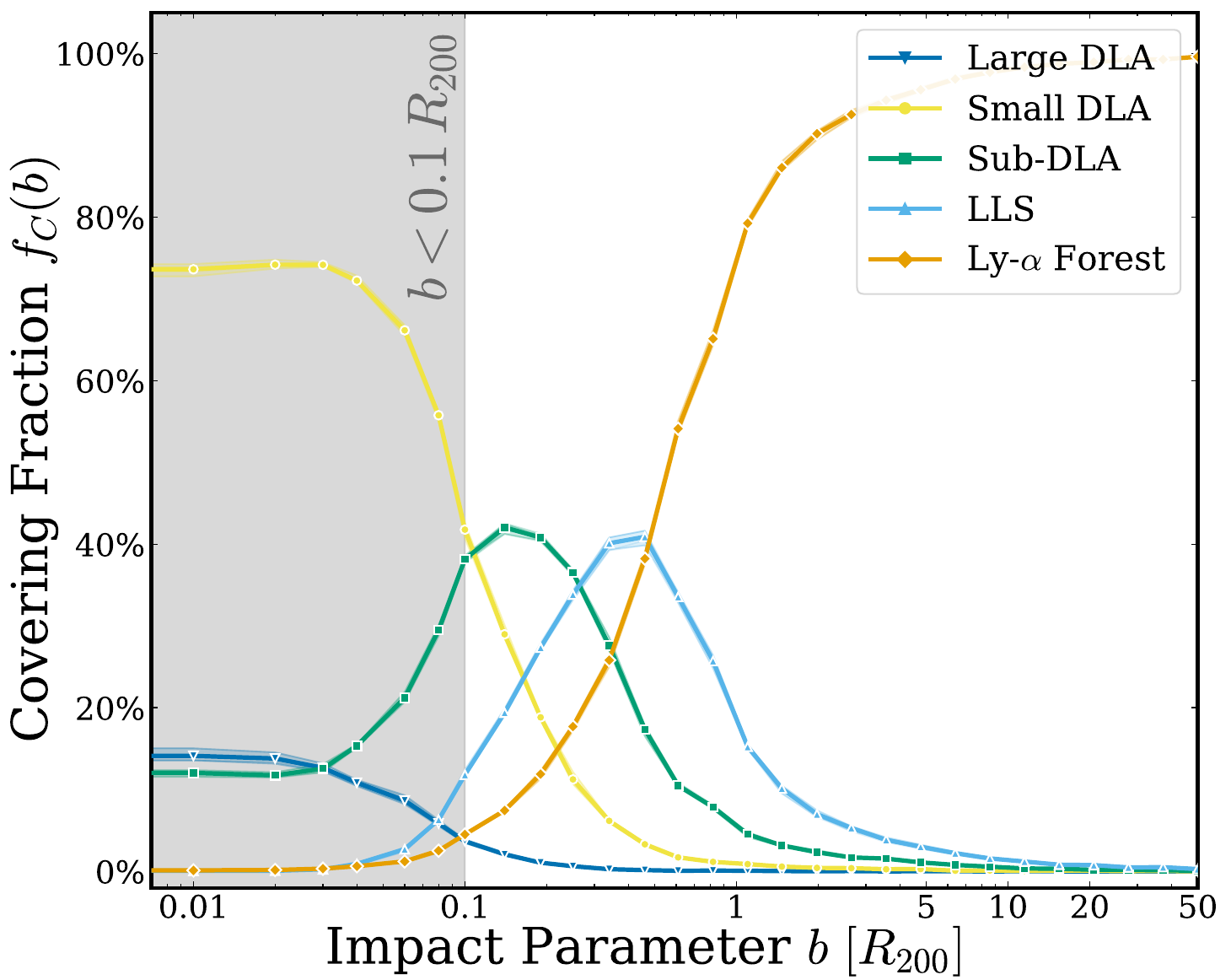}
    \caption{Differential covering fraction $f_C(b)$ in TNG50-2 at $z\simeq3$,
    for the full halo sample. Each curve gives the fraction of sightlines whose
    \texttt{HCD\_only} column density falls in a given class, as a function of
    impact parameter $b$ in units of \Rvir. The five classes are mutually
    exclusive and sum to unity at each $b$. Bands are the $\pm3\sigma$ scatter
    over five halo-sample realizations. The grey region marks impact parameters
    below the resolution floor of Sect.~\ref{subsec:halo_sample_skewers}.}
    \label{fig:hcds_percent_vs_b}
\end{figure}

Figure~\ref{fig:hcds_percent_vs_b} shows the decomposition for the full halo
sample. At the smallest separations, $b\lesssim0.04\,\Rvir$, the small-\dla\
fraction is flat at $\simeq74\%$, with large \dlas\ and sub-\dlas\ accounting for
$\simeq14\%$ and $\simeq12\%$ respectively and crossing each other near
$b\approx0.03\,\Rvir$; the \lls\ and forest channels are empty to within the
realization scatter, so the inner halo is covered by damped absorption along
nearly every line of sight. Large \dlas\ are confined to this region: their
fraction falls below $5\%$ at $b\approx0.1\,\Rvir$ and is negligible beyond
$\approx0.2\,\Rvir$, so the class plays no role in the radial sequence described
below.

These separations lie below the resolution floor of Sect.~\ref{subsec:halo_sample_skewers}, and the plateau values are quoted as indicative. Fig.~\ref{fig:app_covering_resolution} in Appendix~\ref{subsec:app_resolution_covering} shows that the split of the damped population between the small- and large-\dla\ classes is the quantity most sensitive to resolution. At $b\lesssim0.1\,\Rvir$, the large-\dla\ fraction varies significantly across the resolution ladder: it changes from $\simeq 14\%$ in TNG50-2 to $\simeq 29\%$ in TNG50-3, and reaches $\simeq 46\%$ in the lower-resolution TNG50-4 run. This indicates that the large-\dla\ fraction generally drops with increasing resolution. Correspondingly, the small-\dla\ fraction raises from $\simeq 73\%$ in TNG50-2 to $\simeq 63\%$ in TNG50-3 (starting from $\simeq 46\%$ in TNG50-4).
When considering the \dlas\ as a single class, combining the contributions from both subclasses, their relative covering fraction becomes less sensitive to changes in resolution with differences below $\lesssim 5\%$ even within the $b\leq 0.1\,\Rvir$ regime.
Above that separation, all the curves from different resolution levels agree to within $\lesssim 10\%$ and the qualitative trends remain persistent.

Outside the impact parameter resolution floor the composition reorganizes over less than a decade in impact
parameter, and it does so entirely within the virial radius. The three remaining
classes cross in order of decreasing column density: small \dlas\ are overtaken
by sub-\dlas\ at $b\approx0.11\,\Rvir$, sub-\dlas\ by \llss\ at
$b\approx0.3\,\Rvir$, and \llss\ by the forest at $b\approx0.5\,\Rvir$. The first
of these sits at the resolution floor and is correspondingly the least secure.
Each \hcd\ class peaks in a different radial range, the sub-\dla\ fraction at
$b\approx0.14\,\Rvir$ with $f_C\approx0.42$ and the \lls\ fraction at
$b\approx0.35\,\Rvir$ with $f_C\approx0.41$; no class other than the forest
reaches half the sightlines at any separation. Between $0.1$ and $0.5\,\Rvir$ the
population is mixed, with different classes contributing comparably. This
sequence traces the declining projected \HI\ surface density: each class is
sampled preferentially at the radius where the surface density matches its
column-density interval.

The forest fraction rises monotonically but does not saturate at the halo
boundary: it reaches $\simeq78\%$ at $b=\Rvir$, and
approaches unity only beyond $\approx20\,\Rvir$. This slow approach is a
consequence of the aperture defined in Sect.~\ref{subsec:halo_sample_skewers}: a
sightline passing several \Rvir\ from a halo still associates gas within
$\pm5\,\Rvir$ of the halo centre along the line of sight, so what remains in the
\texttt{HCD\_only} channel at large $b$ is material selected by proximity to the
halo along the sightline --- satellites, filaments and neighbouring halos ---
rather than chance intersections with unrelated structure. Appendix~\ref{app:aperture}
verifies this directly: removing the aperture leaves the radial ordering and the
crossing scales unchanged, and only
the large-$b$ tail is affected, where every class then approaches the cosmic
mean covering fraction of a random sightline through the box.

\subsubsection{Dependence on host-halo mass}
\label{subsubsec:mass_dependence}

\begin{figure}
    \centering
    \includegraphics[width=\columnwidth]{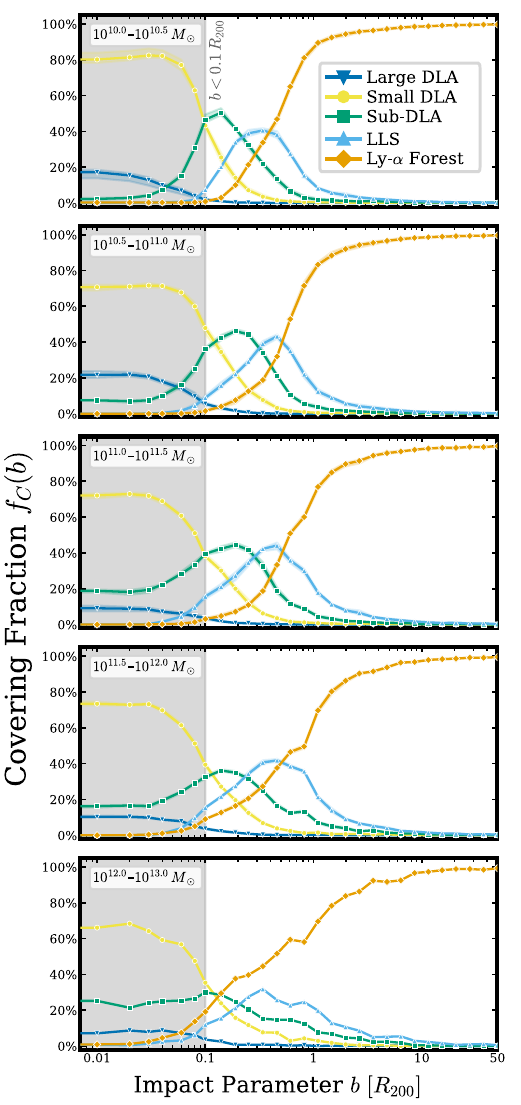}
    \caption{Differential covering fraction $f_C(b)$ in TNG50-2 at $z\simeq3$,
    decomposed by host-halo mass. Each panel shows one of the five mass bins of
    Sect.~\ref{subsec:halo_sample_skewers}, labelled at the top left, with
    classes as in Fig.~\ref{fig:hcds_percent_vs_b}. Bands
    are the $\pm3\sigma$ scatter over five halo-sample realizations. The five
    classes are mutually exclusive and sum to unity at each $b$. Grey regions
    mark impact parameters below the resolution floor of
    Sect.~\ref{subsec:halo_sample_skewers}.}
    \label{fig:covering_mass_split}
\end{figure}

The sample of Sect.~\ref{subsec:halo_sample_skewers} spans three decades in halo
mass, so the combined decomposition of Fig.~\ref{fig:hcds_percent_vs_b} averages
over very different gas environments. Figure~\ref{fig:covering_mass_split}
repeats the measurement in each mass bin, with $1800$ skewers per impact
parameter in the four lower bins and $444$ in the
$10^{12.0}$--$10^{13.0}\,\Msun$ bin.

The decomposition is largely insensitive to halo mass. The radial ordering of the
classes, the impact parameters at which they peak and the radius at which the
forest reclaims half the sightlines are consistent across the five bins, so the
organization described in Sect.~\ref{subsubsec:covering_fraction} is a property of
the halo population as a whole. Since impact
parameters are measured in units of \Rvir, this means the neutral gas
distribution around a halo is mostly set by its virial radius over the mass range we
sample.

Two departures are visible. First, the sequence becomes less sharply defined
towards high mass: the sub-\dla\ and \lls\ peaks weaken from $f_C\simeq0.50$ and
$0.45$ in the lowest bin to $f_C\simeq0.28$ and $0.37$ in the highest, and the
transition between classes spreads over a wider range in $b$. Second, inside the
$b\lesssim0.1\,\Rvir$ plateau the internal ordering of the sub-\dla\ and
large-\dla\ fractions varies with mass; we do not interpret this ordering, since
Appendix~\ref{subsec:app_resolution_covering} (Fig.~\ref{fig:app_covering_mass_resolution})
shows it is not stable across the resolution ladder. Above $b\gtrsim0.1\,\Rvir$,
the same comparison shows differences between resolution levels remain below
$\lesssim10\%$ (a conservative upper bound) and the trends discussed here are stable across all mass bins.

\subsubsection{Column-density distributions across impact parameter}

\begin{figure}
    \centering
    \includegraphics[width=\columnwidth]{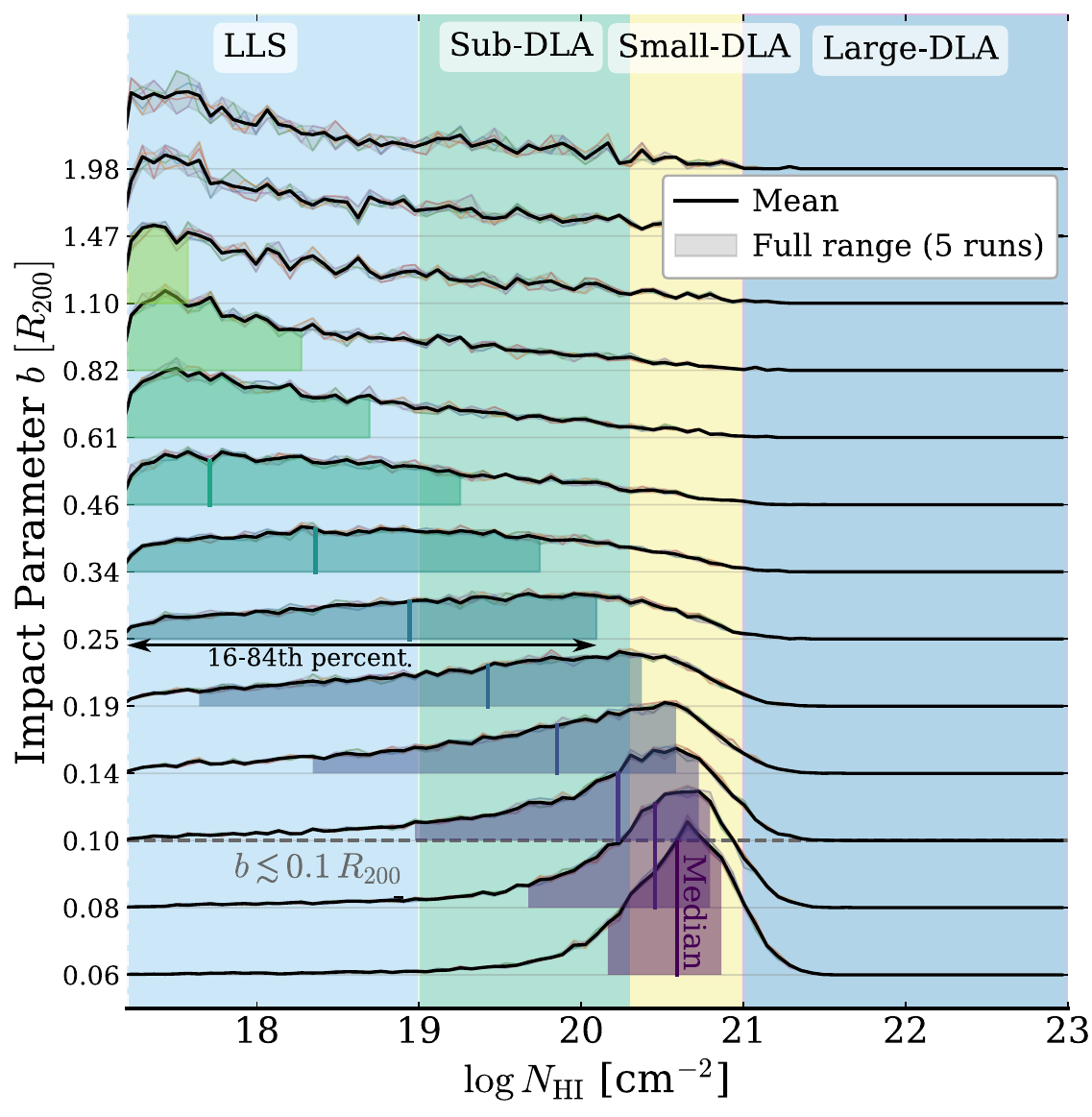}
    \caption{Distribution of the skewer-level \texttt{HCD\_only} column density
    as a function of impact parameter, in TNG50-2 at $z\simeq3$ for the full
    halo sample. Each curve is a normalized distribution at the impact parameter
    given on the ordinate, built from the sightlines that contain an \hcd\ at
    that $b$, and is offset vertically for clarity. Vertical ticks mark the
    median of each distribution and shaded intervals the $16$th--$84$th
    percentile range, both coloured by the median value. Background shading
    marks the absorber classes of Sect.~\ref{subsubsec:covering_fraction}.
    Bands are the full range over five halo-sample realizations. The dashed line
    marks the resolution floor of Sect.~\ref{subsec:halo_sample_skewers}; curves
    below it are shown but not interpreted.}
    \label{fig:nhi_ridge_allmasses}
\end{figure}

The covering fraction assigns each sightline to one of five classes, so it does
not record where inside a class a sightline falls, and the fractions it reports
depend on the boundaries chosen. Figure~\ref{fig:nhi_ridge_allmasses} shows the
distribution of \NHI\ itself as a function of impact parameter, over
$0.06 \leq b/\Rvir \leq 1.98$, the range over which the composition reorganizes
(Sect.~\ref{subsubsec:covering_fraction}).

Each curve uses only the sightlines that contain an \hcd, and is normalized to
their number; the complementary fraction, those with no \hcd\ at all, is the
forest covering fraction of Fig.~\ref{fig:hcds_percent_vs_b}. The two together
determine the incidence of \hcd\ absorbers at each impact parameter. Medians and
$16$th--$84$th percentile intervals are measured on the sightlines and are marked on the figure.

At the innermost impact parameter shown, $b = 0.06\,\Rvir$, the distribution is
narrow and peaks in the small-\dla\ range, with median $\log\NHI \simeq 20.7$ and
width $\simeq0.8$~dex. This narrow distribution is what produces the flat class
fractions reported inside $0.1\,\Rvir$ in
Sect.~\ref{subsubsec:covering_fraction}. Further out the distribution moves to
lower column densities and broadens. At $b = 0.14\,\Rvir$ the median has dropped
to $\simeq19.7$, within the sub-\dla\ regime, where its covering fraction peaks
(Fig.~\ref{fig:hcds_percent_vs_b}), and the width has grown to $\simeq2$~dex, so
the central $68\%$ of the sightlines spread over the \lls, sub-\dla\ and
small-\dla\ ranges; this is the mixed regime seen between $0.1$ and
${\sim}0.5\,\Rvir$ in Fig.~\ref{fig:hcds_percent_vs_b}. Beyond this point less than $50\%$ of the
sightlines no longer contain an \hcd, and those that do remain broadly
distributed, with \llss\ dominating the \hcd\ population through the low-end tail
close to the forest threshold.

The resolution dependence of this picture is not uniform: it is confined to
$b\lesssim0.1\,\Rvir$, and the three levels agree closely outside it.
Appendix~\ref{subsec:app_resolution_nhi} repeats this measurement at the three
TNG50 levels (Fig.~\ref{fig:app_nhi_resolution}). Below $b\simeq0.1\,\Rvir$ the
peak of the distribution moves with resolution: from
$\log\NHI\simeq21.0$, at the small-/large-\dla\ boundary, in TNG50-4, to
$\log\NHI\simeq20.7$, clearly within the small-\dla\ class, in TNG50-2 --- a
shift of $\simeq0.3$~dex to lower column density with increasing
resolution. This is the origin of the large-/small-\dla\ sensitivity reported in
Sect.~\ref{subsubsec:covering_fraction}: coarser resolution pushes part of the
innermost population across the small-/large-\dla\ boundary. Above
$b\simeq0.1\,\Rvir$ the three levels overlap closely, and the radial migration
described above (the fall in median \NHI\ and the broadening of the
distribution between $0.05$ and $2\,\Rvir$) is reproduced at every
resolution.

\subsection{Voigt-profile characterization}
\label{subsec:voigt_diagnostics}

Sect.~\ref{subsec:nhi_statistics} measures the total column density of each
sightline, but not the shape of the absorption that produces it. Two systems with
the same \NHI\ remove different amounts of transmission from a spectrum depending
on how that absorption is distributed in velocity. Observational analyses
characterize \hcds\ as a single absorbing element with one column density and one
width, the description assumed whenever a Voigt profile is fitted to a system in
order to mask it. We measure how faithfully it reproduces the systems our
pipeline isolates, which is a prerequisite for any prescription that assigns
\hcd\ absorption to halos.

We fit a single Voigt component to each retained system
(Sect.~\ref{subsec:halo_sample_skewers}) and compare the fitted column density
with the one measured on the cells, which the simulation gives exactly. The
residuals bound the accuracy the description can reach, and their structure
identifies the configurations that break it.

\begin{figure}
    \centering
    \includegraphics[width=\columnwidth]{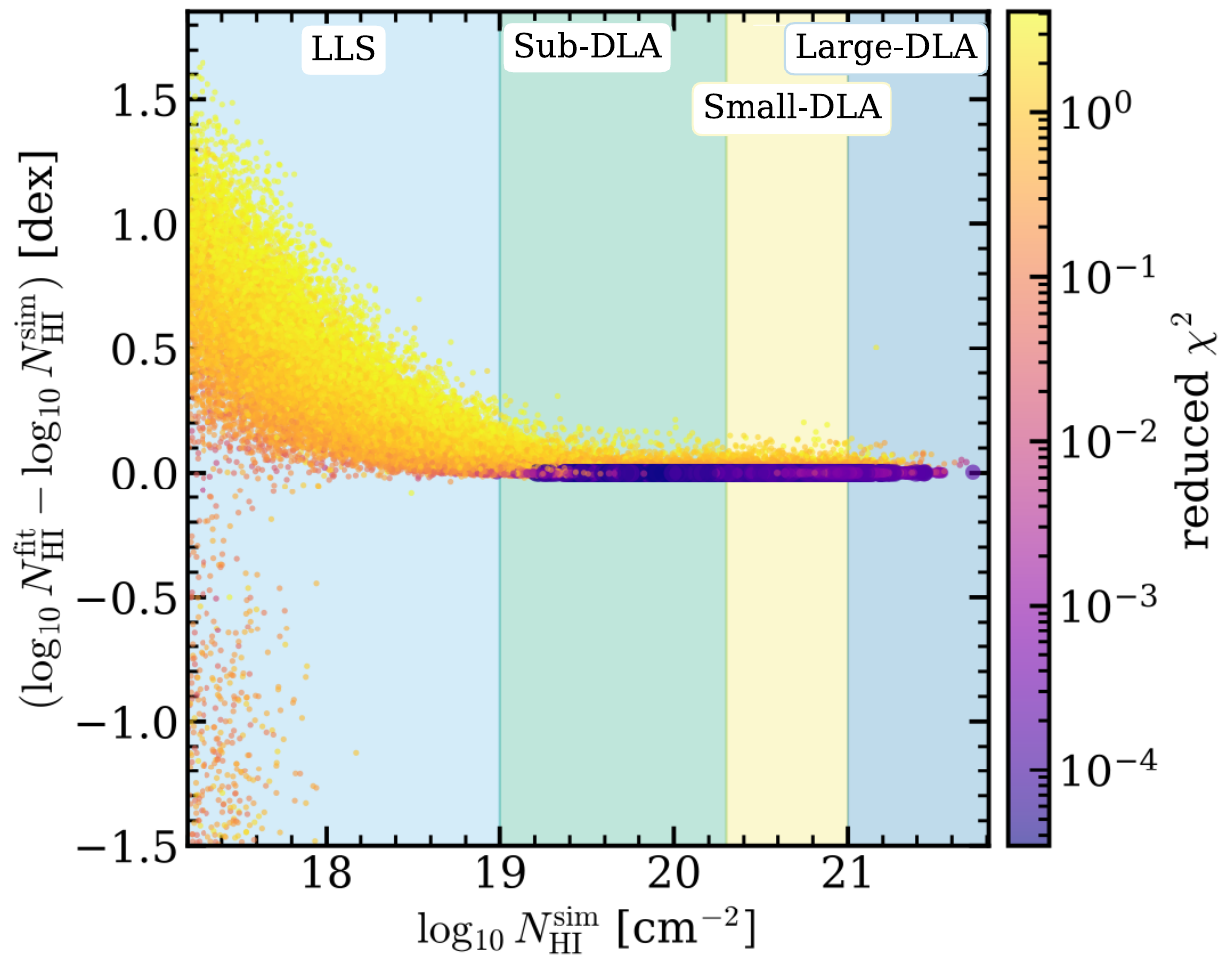}
    \caption{Column-density residual of the single-component Voigt fit,
    $\Delta\log\NHI = \log\NHI^{\rm fit} - \log\NHI^{\rm sim}$, against the column
    density measured on the cells, in TNG50-2 at $z\simeq3$. Point colour gives
    the reduced $\chi^2$ of the fit and marker area scales with $1/\chi^2_\nu$,
    so well-fitted systems are the large dark points. The dashed line marks
    perfect recovery and the background shading the absorber classes of
    Sect.~\ref{subsubsec:covering_fraction}. Systems with $\log\NHI\gtrsim19$ lie
    on the dashed line at every $\chi^2_\nu$; the \lls\ class is offset to
    positive $\Delta\log\NHI$ even where $\chi^2_\nu$ is small.
    }
    \label{fig:residual_NHI_fit_sim}
\end{figure}

Figure~\ref{fig:residual_NHI_fit_sim} shows the residual against the measured
column density. The scatter narrows with increasing \NHI\ and collapses onto the
identity line above $\log\NHI\simeq19$, roughly the sub-\dla\ boundary, where
fitted and measured column densities agree for essentially every system: the
damping wings fix \NHI\ whatever the rest of the profile looks like. Below that
value the residuals spread over more than a decade and are asymmetric. Most \lls\
fits overshoot the measured column density, with a sparser group falling below
it. These residuals correlate with fit quality, the overshoot tail being
populated almost entirely by high $\chi^2_\nu$ systems while the fits near the
identity line are the well-fitted ones, so a single component fails to describe
the absorption of these systems and returns the wrong column density in the same
cases.

To identify what it fails to reproduce, Fig.~\ref{fig:voigt_gallery} presents the
same fits against $\chi^2_\nu$ rather than column density, separated by class,
together with the profiles of representative systems. Three populations are
visible in the top panel. The sub-\dlas, small \dlas\ and large \dlas\ occupy a
horizontal locus at $\Delta\log\NHI\simeq0$ spanning several decades in
$\chi^2_\nu$, so position along it measures profile quality alone and says
nothing about \NHI. The \llss\ form the other two populations, split by the sign
of the residual into an overshooting and an undershooting group, which
correspond to two distinct configurations. The panels analyze different cases,
and for two of them Appendix~\ref{app:cell_level} shows the cell-level
decomposition of Sect.~\ref{subsec:hcd_isolation}, where the origin of the
departure is visible directly in the column-density structure along the
sightline:

\begin{figure}
    \centering
    \includegraphics[width=1.0\columnwidth]{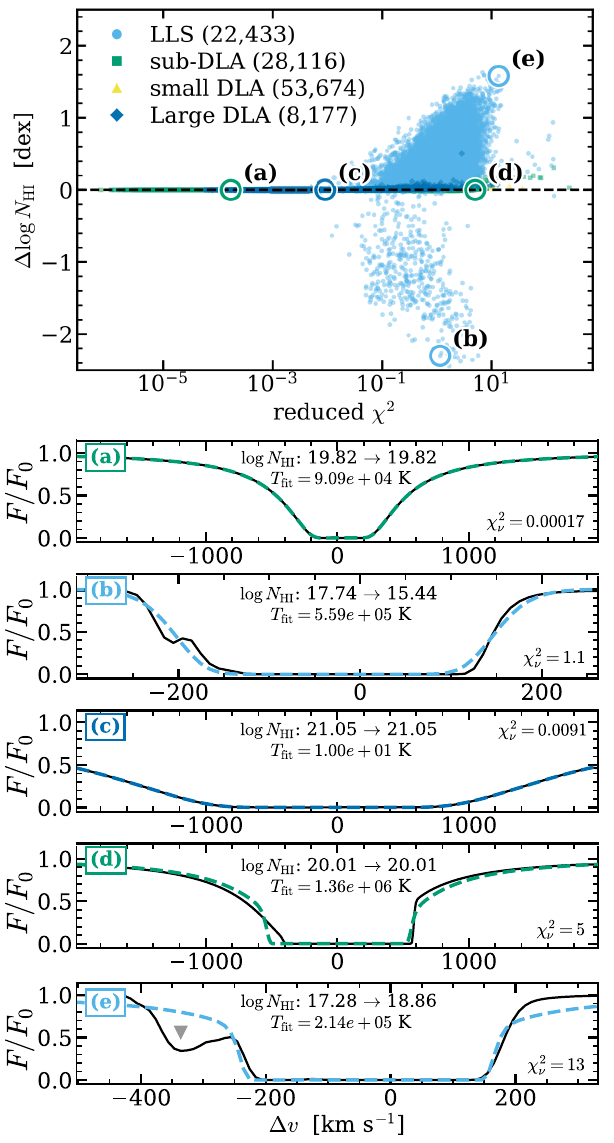}
\caption{Success and failure example modes of the single-component Voigt description. \emph{Top:}
every fitted \hcd\ in the $(\chi^2_\nu,\Delta\log\NHI)$ plane, coloured and
marker-coded by absorber class; the dashed line marks perfect recovery.
\emph{Panels:} simulated \texttt{HCD\_only} flux (black) and best-fit profile
(dashed) for the six lettered systems, analysed individually in
Sect.~\ref{subsec:voigt_diagnostics}.
}
    \label{fig:voigt_gallery}
\end{figure}
\begin{itemize}

\item \textbf{(a) The reference case.} A sub-\dla\ recovered at
$\chi^2_\nu\sim10^{-4}$ and $\Delta\log\NHI=0.00$. The fitted and simulated
profiles are indistinguishable, with no discernible substructure: one absorbing
element with one column density reproduces the absorption
completely. This is the behaviour the single-component description assumes.

\item \textbf{(b) Saturated with substructure: the undershoot cloud.} An \lls\
saturated over $\sim350\;\kms$, whose measured profile from the simulation shows a step on the
negative-velocity side where the flux levels off near $F/F_0\simeq0.4$ before
falling into the trough. The fitted component has no such feature: it reproduces
the trough and the outer flanks at $\chi^2_\nu=1.1$, but returns a column density $2.30$~dex below the measured one. The
cell-level view (Appendix~\ref{app:cell_level}) shows the step comes from a
weaker column-density structure adjacent to the main peak in velocity space and
retained in the same basin. A single component cannot represent both, and
compensates by narrowing the profile and lowering \NHI. This is the
configuration behind the lower \lls.

\item \textbf{(c) Correct \NHI, profile truncated by the box.} A large \dla,
recovered to $0.00$~dex because the damping wings pin the column density even
with the core fully saturated. The wings extend over more than $10^{3}\;\kms$
and wrap around the periodic velocity axis, so the absorption appears at both
ends of the skewer.

\item \textbf{(d) Correct \NHI, wrong profile.} A sub-\dla\ recovered to
$0.00$~dex, yet at $\chi^2_\nu=5$. The core is reproduced while the wings are
systematically offset from the simulated flux. The cell-level view
(Appendix~\ref{app:cell_level}) identifies the cause: a single \hcd\ peak
carries essentially all of the column, but several forest-level structures (orders of magnitude weaker in column density terms) lie on the sides of it in velocity and are retained in the same basin, so the absorption assigned to the system is suppressed over a wider interval than a single Voigt profile of that column density can
reproduce. The total column is recovered exactly because it is set by the
central peak, while the shape is not.

\item \textbf{(e) Two systems in one basin: the overshoot cloud.} An \lls\ with
a second absorption near $\Delta v\simeq-350\;\kms$ (grey triangle), retained in
the same basin by the segmentation of Sect.~\ref{subsec:hcd_isolation}. The
single component broadens to cover both and returns a column density $1.58$~dex
above the measured one, at $\chi^2_\nu=13$. The configuration is that of
panel~(b) with a stronger and more widely separated companion, and it produces
the opposite bias: where~(b) narrows and undershoots, (e) broadens and
overshoots. This is the configuration behind the upper \lls\ cloud.
\end{itemize}

The cases separate along two largely independent axes. Whether \NHI\ is recovered
depends on what the damping wings pin down. On the damped branch the total
optical depth is, to good approximation, insensitive to how the column is
distributed in velocity, so two blended components with columns $N_1$ and $N_2$
leave nearly the same imprint as one component with $N_1+N_2$ and a single fit
recovers the sum even when it cannot recover the shape. That is the situation in
panel~(d), and it holds trivially in panels~(a), and (c), where the correct
total and the correct shape coincide because there is only one structure to
begin with. Within the \lls\ class this
compensation does not operate, and the fitted \NHI\ misses the true value by more
than a decade in either direction, as in panels~(b) and~(e). Whether the profile
is reproduced is set instead by how many structures the sightline contains and
how the absorption is distributed among them, which is the multi-component
character that panel~(d) shows to be the exception among the damped classes and
panels~(b) and~(e) show to be the norm among \llss. A prescription that assigns
one absorber per system is therefore adequate for the column density of the
damped and sub-damped population, and inadequate for the \lls\ class.

\section{Discussion and Conclusions}\label{sec:discussion}

We have characterized the imprint of high column density systems on \lya\ skewers
as a function of the properties of the halos that host them. The measurement
rests on three elements: a skewer-level procedure that identifies
high-column-density structures on the gas cells themselves, before the optical
depth is computed, and returns a \texttt{HCD\_only} and a \texttt{Forest\_only}
spectrum for each sightline; a sampling campaign that pierces halos of known mass
at controlled distances, so that the absorption recorded can be read directly as
a function of halo mass and halo-centric distance; and the TNG50 resolution
ladder, which fixes the baryonic model and isolates numerical resolution as a
systematic. Three results follow.

\emph{The absorber population is organized radially, and the organization is
complete within the virial radius.} The classes cross in order of decreasing
column density: small \dlas\ are overtaken by sub-\dlas\ at $b\approx0.11\,\Rvir$,
sub-\dlas\ by \llss\ at $0.3\,\Rvir$, and \llss\ by the forest at $0.5\,\Rvir$.
The core is covered by damped absorption almost without exception, and the forest
reclaims the majority of sightlines only outside \Rvir, reaching $\simeq78\%$
there and approaching unity beyond $\approx20\,\Rvir$. The class decomposition is
a coarse-graining of a single \NHI\ distribution that moves to lower columns and
broadens with radius, so the peaks in the covering fraction are not distinct
populations but the radii at which one declining profile passes through each
interval.

\emph{Expressed in units of \Rvir, this sequence is nearly independent of halo
mass.} Over three decades in \Mvir\ the class ordering, the peak positions and the
radius at which the forest takes over are consistent, so the virial radius sets
the scale of the neutral gas distribution over the range we sample. The one
systematic trend is a weakening and broadening of the sequence towards high mass.

\emph{The single-absorber description splits at the sub-\dla\ boundary.} Above it
the damping wings fix the column density, which is recovered for essentially every
system. Below it, the fitted column density residuals for \llss\ are spread over more than a decade, driven by systems that contain more than one
structure rather than one. The failure is not one of fitting: the systems are simply not single absorbing
elements, and no choice of profile parameters can represent absorption produced
by several structures at once.

Two applications follow, one in each direction. The first is direct, from halos to spectra: for mock generation, \hcds\ drawn from a global column-density distribution and assigned an effective bias \citep[e.g.,][]{2026PhRvD.113b3520C, 2026arXiv260727412R} could instead follow a prescription conditioned on the halo, in which the incidence and strength of an absorber are set by the mass of the nearest halo and the transverse distance to it. Our measurements supply what such a prescription needs and constrain its form, while the \texttt{HCD\_only}/\texttt{Forest\_only} decomposition provides the test the resulting mocks must pass. The second is the inverse, from spectra to halos: because the absorber a sightline records depends on the mass of the nearest halo and the transverse distance to it, the \hcds\ observed in a spectrum carry information about the halo population they trace, which our results quantify.

Three extensions also follow directly from the framework as it stands. The first is in resolution: large \dlas\ reach an appreciable covering fraction only at $b\lesssim0.1\,\Rvir$, below the resolution floor of the campaign, and their fraction there is the least converged quantity we measure (Appendix~\ref{subsec:app_resolution_covering}), so pinning it down calls for higher-resolution runs. The second is in redshift: \lya\ forest analyses draw on $2\lesssim z\lesssim4$, across which the neutral gas content of halos evolves substantially, so the radial organization measured here has to be established over the full interval before it can inform mocks spanning it. The third is in the baryonic model, which the TNG50 ladder holds fixed by construction; the zoom-in re-simulations of \citet{2026arXiv260713151M} follow the same halos under varied feedback prescriptions, and applying the pipeline to them would separate the dependence on halo properties reported here from the dependence on the subgrid physics that sets how much neutral gas the circumgalactic medium retains.

\begin{acknowledgments}

We thank Andreu Font-Ribera for many useful discussions throughout this work.
This paper builds on the master's thesis of VRS, and we are grateful to Simeon
Bird and Ignasi P\'erez-R\`afols, members of its examining committee, for their
comments at that stage and on the elaboration of this work. We also thank Matteo
Zennaro, Joaqu\'in Armijo, and M. F. Ruiz-Herrera Bernal for helpful discussions.

VRS is supported by the S\~ao Paulo Research Foundation (FAPESP) via a Master's
fellowship (Process No.~2023/18127-9).

DLC is supported by the S\~ao Paulo Research Foundation (FAPESP) via a
Postdoctoral Fellowship (Process No.~2024/05768-9), linked to the FAPESP Thematic
Project (Process No.~2019/26492-3), and by a FAPESP Research Internship Abroad
Fellowship (BEPE, Process No.~2025/24623-4).

RA acknowledges support from the S\~ao Paulo Research Foundation (FAPESP) through
the same Thematic Project (Process No.~2019/26492-3) and from the Brazil--France
FAPESP--ANR programme (FAPESP Process No.~2022/03426-8).

JCM acknowledges
financial support from the Spanish Ministry of Science and Innovation (MICINN) through the
Spanish State Research Agency, under Severo Ochoa Centres of Excellence Programme 2025-2029 (CEX2024-001441-S) and the European Union (ERC Consolidator Grant, COSMO-LYA, grant agreement 101044612).

FM acknowledges support by the Simons Collaboration
on “Learning the Universe”.

The IllustrisTNG simulations are publicly available at \url{https://www.tng-project.org} \citep{2019ComAC...6....2N}.

Generative artificial-intelligence tools, in particular Claude (Anthropic), were used to assist with writing short pieces of code and resolving numerical issues, and marginally for proofreading. All content was reviewed and verified by the authors.
\end{acknowledgments}

\begin{contribution}

Contributions follow the CRediT taxonomy (Contributor Roles Taxonomy;
ANSI/NISO~Z39.104-2022).

\begin{description}
\item[VRS] Conceptualization, methodology, software, formal analysis,
investigation, visualization, writing --- original draft, review and editing.
\item[DLC] Conceptualization, methodology, validation, supervision, writing ---
original draft, review and editing.
\item[RA] Conceptualization, funding acquisition, supervision, writing --- review
and editing.
\item[JCM] Methodology, writing --- review and editing.
\item[FM] Resources, software, writing --- review and editing.
\end{description}

All authors reviewed and approved the final manuscript.
\end{contribution}

\software{
\textsc{Python},
\textsc{NumPy} \citep{harris2020array},
\textsc{SciPy} \citep{2020SciPy-NMeth},
\textsc{Matplotlib} \citep{Hunter:2007},
\textsc{JAX} \citep{jax2018github},
\textsc{Astropy} \citep{2022ApJ...935..167A},
\textsc{fake\_spectra} \citep{2017_Bird_theory_Lya_mock_Fake_Spectra}
}

\section*{Data Availability}
The modified version of \texttt{fake\_spectra} used to produce the \texttt{HCD\_only} and \texttt{Forest\_only} decompositions, including the deblending step of Section~\ref{subsec:hcd_isolation} and the optical-depth correction of Section~\ref{subsec:fake_spectra_skewers}, is publicly available at \url{https://github.com/VictorRoberto132/FakeSpectra_DLAs}. The skewer catalogues underlying the figures in this paper will be made available on Zenodo upon publication.
All figures use the Okabe--Ito qualitative colour palette \citep{OkabeIto2008},
chosen to remain legible under the common forms of colour vision deficiency; in
addition, absorber classes are distinguished by line style and marker as well as
by colour, so no figure relies on colour alone.

\appendix

\section{Numerical resolution and convergence}\label{app:resolution_tests}

\hcds\ are regions of high gas density, and in a moving-mesh code such as
\textsc{arepo} these are where the Voronoi cells are smallest and most numerous,
so the absorption we measure is sensitive to the mass resolution of the run. This
appendix establishes the range over which our measurements are converged. We use
the TNG50 ladder, TNG50-4, TNG50-3 and TNG50-2, which share an identical
baryonic model and differ only in the number of resolution elements
(Table~\ref{tab:tng50}), so any difference among them is attributable to
resolution alone. All tests use the halo selection, skewer generation and
analysis of the main text at $z\simeq3$, in matched halo-mass and
impact-parameter bins.

\subsection{Covering fractions}\label{subsec:app_resolution_covering}

\begin{figure}
    \centering
    \includegraphics[width=0.5\textwidth]{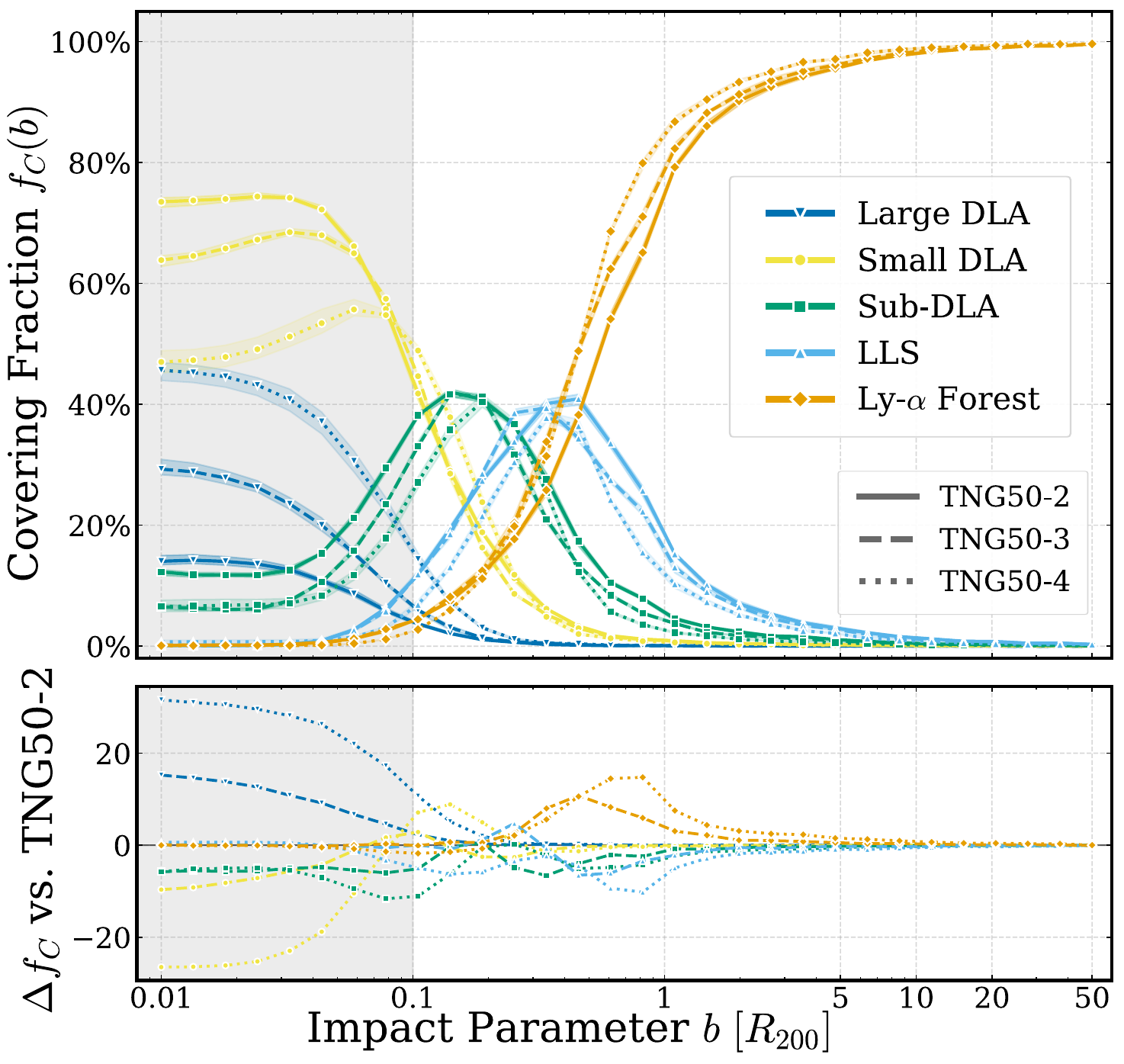}
    \caption{Differential covering fraction $f_C(b)$, as in
    Fig.~\ref{fig:hcds_percent_vs_b}, measured in TNG50-4, TNG50-3 and TNG50-2
    with identical pipeline and halo selection. \emph{Top:} the five classes,
    each at the three resolution levels; line style encodes resolution, color
    encodes class. \emph{Bottom:} difference of each level with respect to
    TNG50-2, in percentage points, isolating the resolution dependence
    directly. Bands are the $\pm3\sigma$ scatter over five halo-sample
    realizations. The shaded region marks $b<0.1\,\Rvir$.
    }
    \label{fig:app_covering_resolution}
\end{figure}

Figure~\ref{fig:app_covering_resolution} isolates the resolution dependence
directly in relation with the results presented in Fig.~\ref{fig:hcds_percent_vs_b}. Below $b\simeq0.1\,\Rvir$ the three levels disagree substantially: at
the smallest impact parameter sampled, the large-\dla\ fraction in TNG50-4
exceeds that of TNG50-2 by $\simeq35$ percentage points, and the
small-\dla\ fraction is lower by a comparable amount, so coarser resolution
shifts part of the innermost population from the small- into the large-\dla\
class, consistent with the peak shift of
Sect.~\ref{subsubsec:covering_fraction}. Because the two offsets are of similar
size and opposite sign, the aggregate damped fraction (small
and large \dlas\ combined) is comparatively stable, differing between TNG50-4
and TNG50-2 by only $\simeq5$ percentage points at the same separation. The same
compensation is visible directly in the column-density distribution
(Fig.~\ref{fig:app_nhi_resolution}), where the peak shifts across the
small-/large-\dla\ boundary with resolution.

Above $b\simeq0.1\,\Rvir$ the three levels agree to within a few percentage
points for every class. The largest residual differences, up to
$\simeq15$ percentage points, appear as a transient feature in the forest fractions around $b\simeq0.3\Rvir$, and fall below
$\simeq10$ percentage points elsewhere. This confirms
the resolution floor adopted in Sect.~\ref{subsec:halo_sample_skewers}: the
classification used throughout Sect.~\ref{subsubsec:covering_fraction} is
resolution-converged for $b\gtrsim0.1\,\Rvir$, and only the internal split of
the damped population inside that radius is not.

\begin{figure*}
    \centering
    \includegraphics[width=1.0\textwidth]{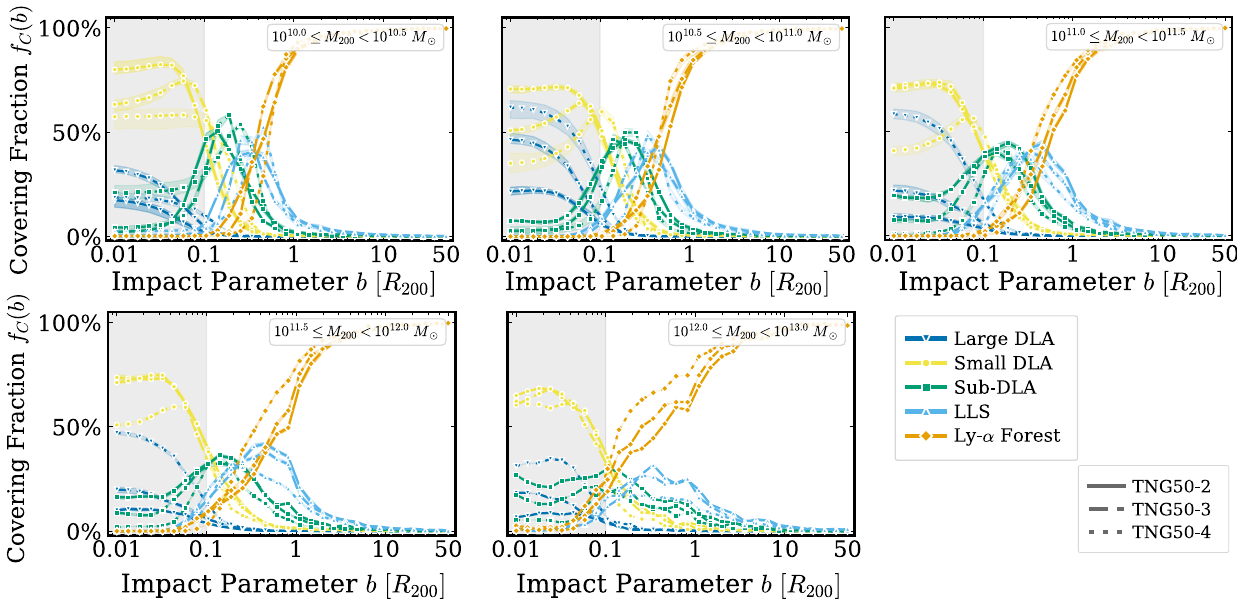}
    \caption{Mass-split covering fractions, as in
    Fig.~\ref{fig:covering_mass_split}, at the three resolution levels.}
    \label{fig:app_covering_mass_resolution}
\end{figure*}

Figure~\ref{fig:app_covering_mass_resolution} repeats the mass split of
Sect.~\ref{subsubsec:mass_dependence} at the three resolution levels. The
core-plateau crossing reported there --- large \dlas\ the second most common
class in the two lowest mass bins, sub-\dlas\ in the three highest --- is a
feature of TNG50-2 alone: at lower resolution the sub-\dla\ fraction inside
$0.1\,\Rvir$ is elevated across all mass bins, encroaching on the region
occupied by large \dlas\ even where TNG50-2 shows a clean separation. The
crossing is therefore not established independently of resolution.

\subsection{Column-density distributions}\label{subsec:app_resolution_nhi}

\begin{figure}
    \centering
    \includegraphics[width=0.7\textwidth]{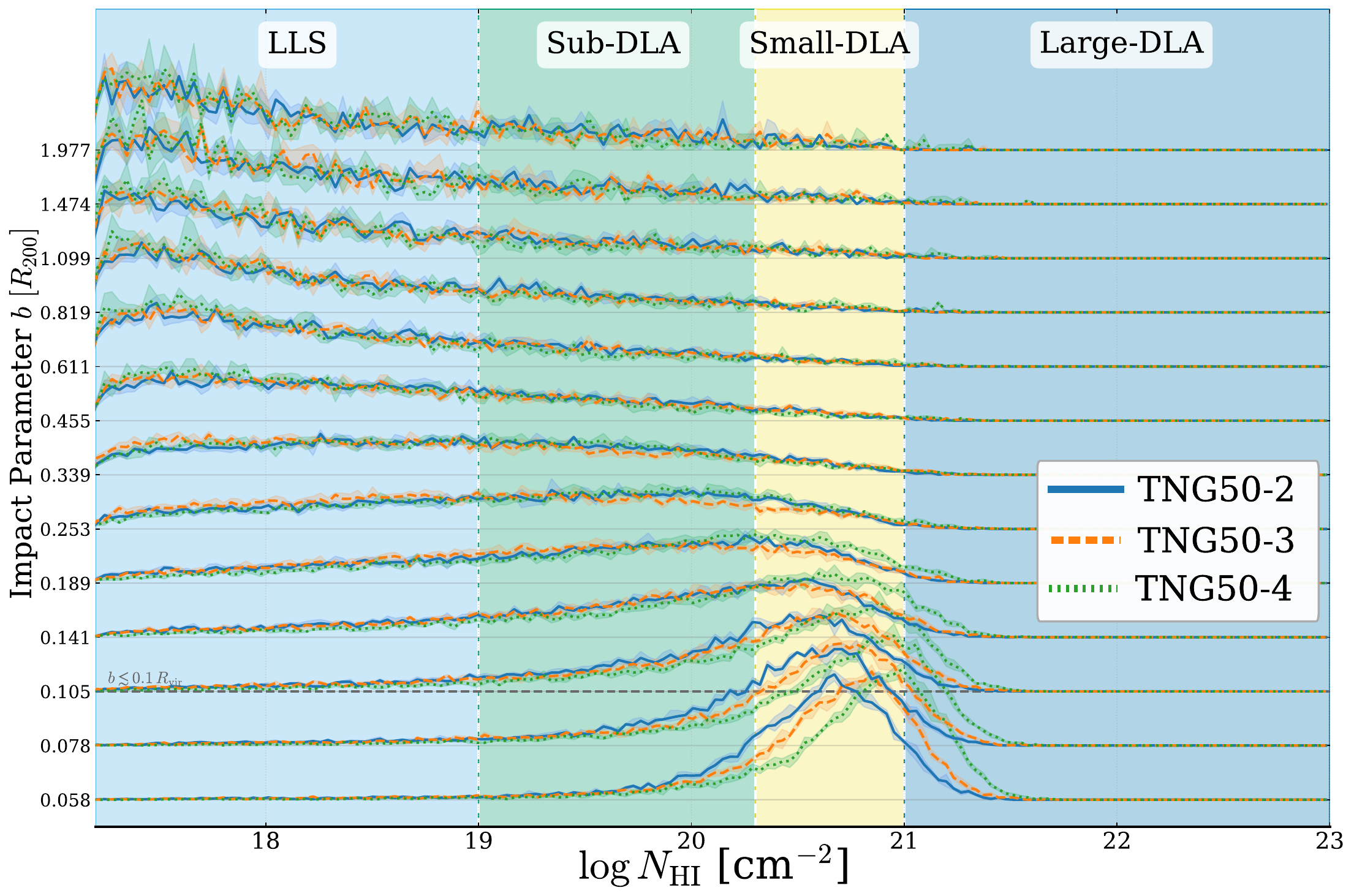}
    \caption{Skewer-level \NHI\ distribution against impact parameter, as in
    Fig.~\ref{fig:nhi_ridge_allmasses}, at the three resolution levels, for a
    fixed set of impact parameters spanning the full sampled range. Background
    shading marks the absorber-class intervals of
    Sect.~\ref{subsubsec:covering_fraction}. Percentages give the forest
    fraction (mean $\pm$ scatter over five halo-sample realizations) at each
    $b$ and resolution level. The shaded region marks $b<0.1\,\Rvir$.}
    \label{fig:app_nhi_resolution}
\end{figure}

Figure~\ref{fig:app_nhi_resolution} shows the equivalent resolution test results to compare against Fig.~\ref{fig:nhi_ridge_allmasses}. For $b\gtrsim0.1\,\Rvir$ the distributions at the three resolution
levels overlap closely: the median, the width and the high-column tail all
agree, and the radial migration of
Sect.~\ref{subsubsec:mass_dependence} --- the distribution moving to lower
\NHI\ and broadening between $0.05$ and $\sim2\,\Rvir$ --- is reproduced at
every level. Below $b\simeq0.1\,\Rvir$ the peak position differs between
levels, from $\log\NHI\simeq21.0$ in TNG50-4 to $\simeq20.7$ in TNG50-2,
a shift of $\simeq{0.3}$~[dex] towards lower column density with increasing
resolution; the forest fractions quoted in the figure are correspondingly
small and consistent with zero within the run-to-run scatter at every level in
this regime. This is the same trend identified independently in the covering
fraction (Fig.~\ref{fig:app_covering_resolution}), measured here directly on
the column-density distribution rather than on the derived class fractions.

\section{The line-of-sight association aperture}\label{app:aperture}

\begin{figure*}
    \centering
    \includegraphics[width=0.8\textwidth]{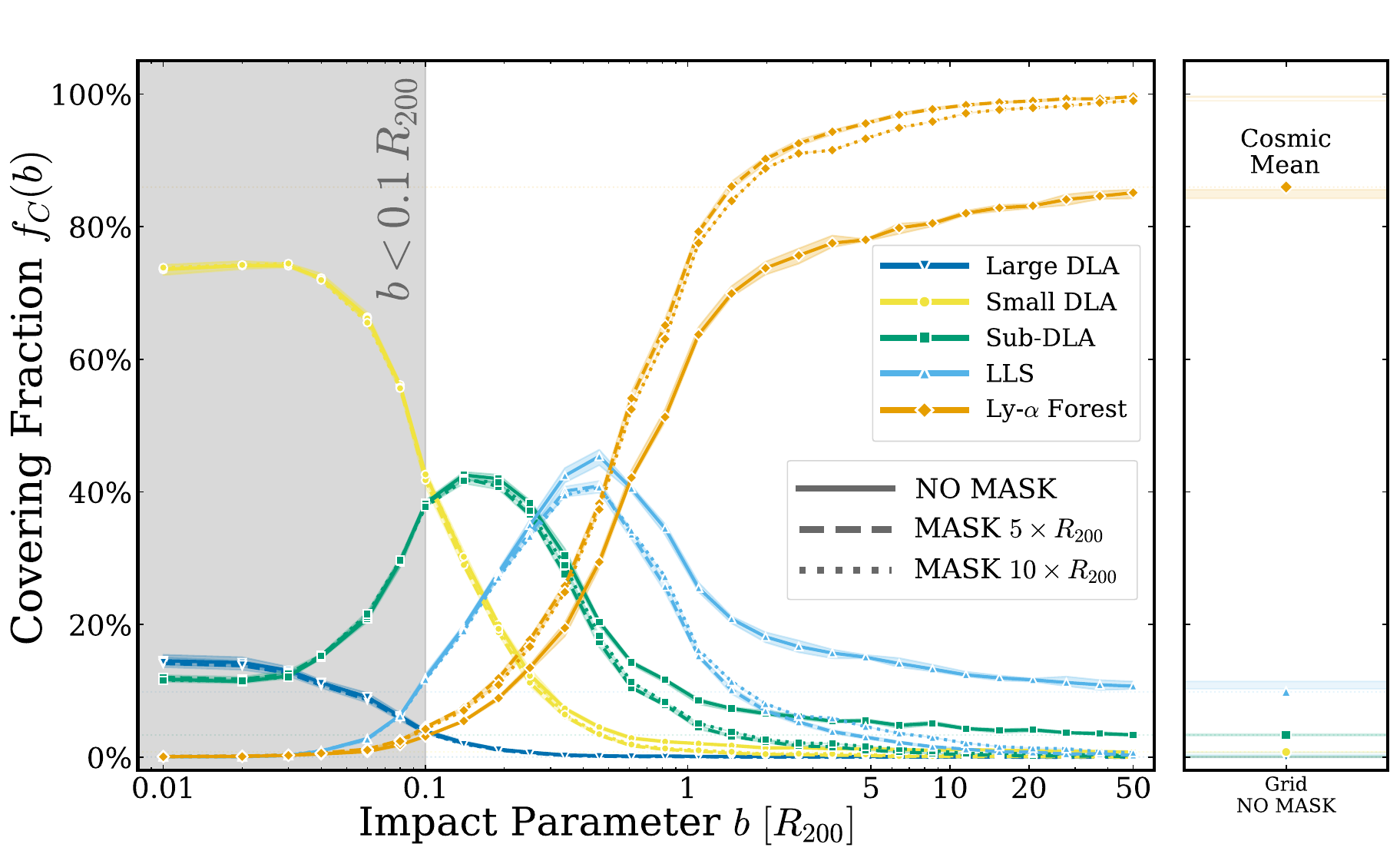}
    \caption{Differential covering fraction $f_C(b)$ in TNG50-2 at $z\simeq3$,
    for three choices of the line-of-sight association aperture of step~4
    (Sect.~\ref{subsec:hcd_isolation}): no aperture (solid), the fiducial
    $\pm5\,\Rvir$ (dashed) and $\pm10\,\Rvir$ (dotted). Without an aperture,
    every basin found anywhere along the periodic sightline is credited to the
    target halo. Bands are the $\pm3\sigma$ scatter over five halo-sample
    realizations and the grey region marks $b<0.1\,\Rvir$, below the resolution
    floor of Sect.~\ref{subsec:halo_sample_skewers}. The right panel gives the
    cosmic mean covering fraction of each class, measured on a grid of
    sightlines placed at random through the box and unrelated to any halo;
    horizontal dotted lines carry those values across the left panel.}
    \label{fig:app_aperture}
\end{figure*}

Figure~\ref{fig:app_aperture} compares the fiducial $\pm5\,\Rvir$ aperture
against $\pm10\,\Rvir$ and against no aperture at all. The three agree closely
over the range where the absorber population reorganizes. Inside
$b\simeq0.5\,\Rvir$ the curves show differences below $5\%$ (largest difference at $b=0.5\,\Rvir$ between \llss\ covering fraction in the ``no mask'' case): at these separations the
sightline passes through gas that lies within a few \Rvir\ of the halo along the
line of sight in any case, so restricting the association changes nothing. The
class crossings of Sect.~\ref{subsubsec:covering_fraction}, and the sub-\dla\
and \lls\ peaks, are therefore independent of this choice.

The curves separate beyond $b\simeq1\,\Rvir$, and they do so in the direction
the aperture is designed to produce. Without an aperture the forest fraction
rises only to $\simeq85\%$ by $b=50\,\Rvir$, and the \lls\ and sub-\dla\
fractions level off near $\simeq12\%$ and $\simeq3\%$ --- precisely the cosmic
mean values measured on random sightlines (right panel). This is the expected
limit: a sightline passing tens of virial radii from a halo still crosses the
rest of the box, and the absorbers it encounters there have no relation to the
target, so an unrestricted association simply recovers the incidence of a
randomly placed sightline. With an aperture in place those absorbers are
assigned to the forest channel, the saturation is removed, and the forest
fraction continues towards unity. The $\pm5\,\Rvir$ and $\pm10\,\Rvir$ apertures
differ from each other by at most a few percentage points at any $b$, so within
this range the result is not sensitive to where exactly the window is placed.

Two conclusions follow. First, the measurements of
Sect.~\ref{subsubsec:covering_fraction} and
Sect.~\ref{subsubsec:mass_dependence}, which focus on $b\lesssim1\,\Rvir$, do not
depend largely on the aperture. Second, the large-$b$ tail reported in
Sect.~\ref{subsec:nhi_statistics} should be read as what the aperture defines it
to be: the incidence of absorbers within $\pm5\,\Rvir$ of the halo along the
sightline, and not the incidence of all structure the sightline happens to
cross, which is the cosmic mean.

\section{Cell-level structure of the departing cases}\label{app:cell_level}

\begin{figure*}
    \centering
    \includegraphics[width=\textwidth]{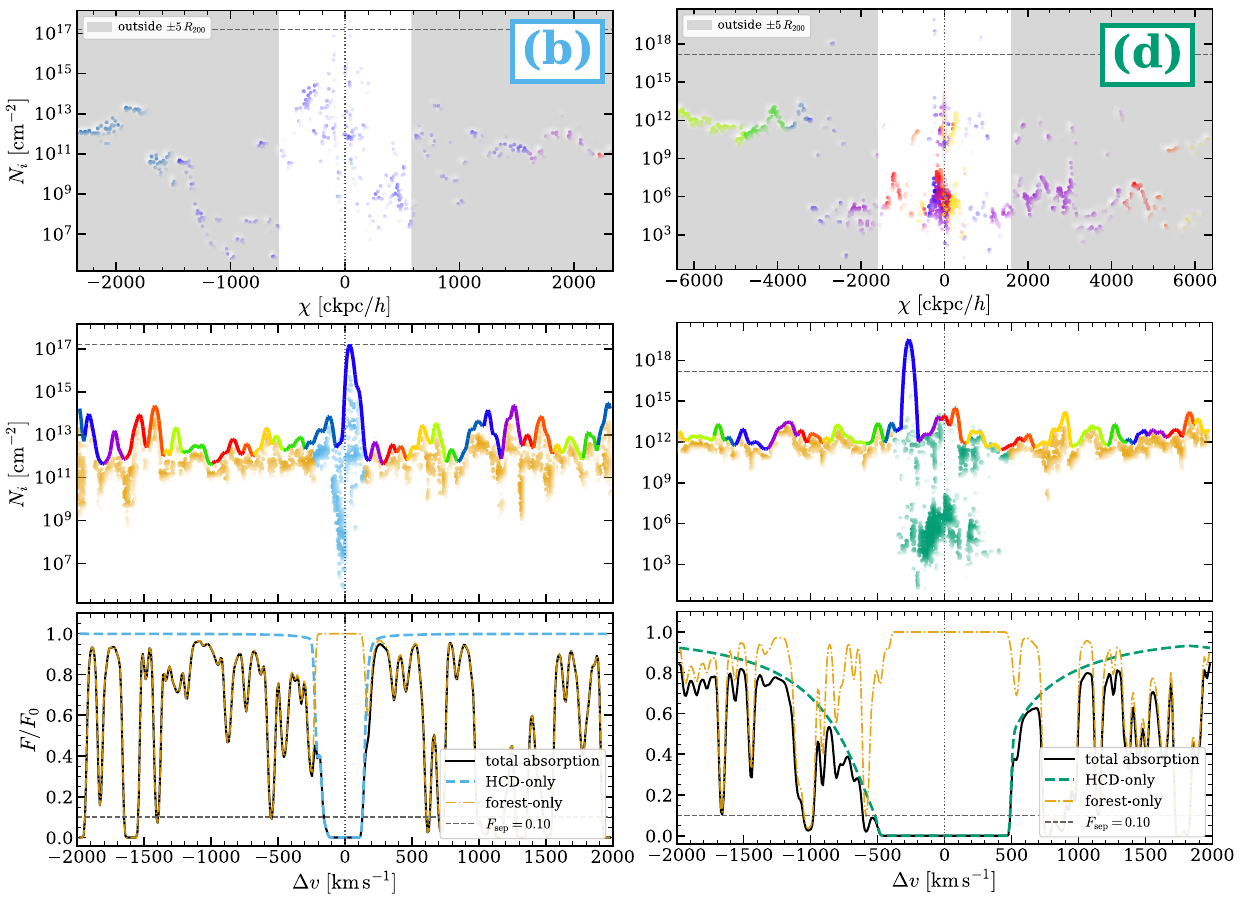}
    \caption{Cell-level decomposition of Sect.~\ref{subsec:hcd_isolation} for
    two of the systems of Fig.~\ref{fig:voigt_gallery}, labelled (b) and~(d) as
    in that figure, whose single-component fits depart from the simulated
    profile in different ways. Panels follow Fig.~\ref{fig:hcd_detection}:
    contributed column per cell against comoving position (top), the same cells
    against velocity together with the smoothed profile and its basins
    (middle), and the resulting total, \texttt{HCD\_only} and
    \texttt{Forest\_only} spectra (bottom). Note the different vertical ranges
    of the two columns.}
    \label{fig:cell_level_bde}
\end{figure*}

Figure~\ref{fig:cell_level_bde} shows the column-density structure behind two of
the departing cases of Sect.~\ref{subsec:voigt_diagnostics}, chosen because they
fail for different reasons.

Case~(b) is an \lls\ whose basin is dominated by a single peak reaching
$\NHI\simeq10^{17}\,\mathrm{cm^{-2}}$, immediately beside which a weaker
structure, roughly an order of magnitude lower in column, is retained in the
same basin by the contrast criterion of step~3. The two are separated by less
than the width of the trough they jointly produce, so the \texttt{HCD\_only}
spectrum in the bottom panel shows a single saturated feature with a shoulder
rather than two resolved lines. A single Voigt component fitted to that profile
cannot reproduce both the flat core and the asymmetric flank, and the fit
settles on a narrower, weaker solution, which is the $2.30$~dex undershoot
reported in Sect.~\ref{subsec:voigt_diagnostics}. The forest channel recovers to
unit transmission immediately outside the trough, confirming that the departure
originates within the basin rather than from absorption assigned elsewhere.

Case~(d) fails in the opposite manner. Its basin contains one narrow peak near
$\Delta v\simeq-300\;\kms$ reaching $\NHI\simeq10^{19}\,\mathrm{cm^{-2}}$, which
carries essentially the entire column of the system and therefore sets the value
a fit recovers. Surrounding it, however, is a broad population of cells at
forest-level columns, $10^{11}$--$10^{13}\,\mathrm{cm^{-2}}$, spread over
several hundred \kms\ on both sides and assigned to the same basin. Individually
negligible, together they suppress the transmission across that whole interval:
the \texttt{HCD\_only} spectrum in the bottom panel remains saturated out to
${\pm}500\;\kms$ and its wings extend well beyond, far wider than the damping
wings of a single absorber of that column density would be. The fit therefore
returns the correct \NHI\ --- fixed by the dominant peak --- while
systematically missing the shape, which is the combination reported as
$\Delta\log\NHI=0.00$ at $\chi^2_\nu=5$.

The two cases bound the ways a basin can depart from a single absorbing element.
In~(b) the departure is a second comparable structure, which biases the fitted
column density. In~(d) it is a continuum of much weaker structures, which leaves
the column density intact and corrupts only the profile. Both are invisible in
the spectrum alone, and both are recovered directly from the cells.

\bibliographystyle{aasjournalv7}
\bibliography{main}{}

\end{document}